\documentclass[fleqn,usenatbib]{mnras}

\usepackage{newtxtext,newtxmath}

\usepackage[T1]{fontenc}
\usepackage{ae,aecompl}

\usepackage{graphicx}	
\usepackage{amsmath}	

\def\kms{{\rm\,km\,s^{-1}}}

\def\kpc{{\rm\,kpc}}

\def\coreNFWtides{{\sc coreNFWtides}}
\def\GravSphere{{\sc GravSphere}}
\def\coreNFW{{\sc coreNFW}}
\def\vsone{v_{s1}}
\def\vstwo{v_{s2}}
\def\vlosfour{\langle v_{\rm los}^4 \rangle}
\def\dd{\text{d}}

\def\={\overline}

\def\lta{\mathrel{\spose{\lower 3pt\hbox{$\mathchar"218$}}
     \raise 2.0pt\hbox{$\mathchar"13C$}}}
\def\gta{\mathrel{\spose{\lower 3pt\hbox{$\mathchar"218$}}
     \raise 2.0pt\hbox{$\mathchar"13E$}}}

\def\sun{\odot}

\def\jcap{JCAP}
\def\lta{\mathrel{\spose{\lower 3pt\hbox{$\mathchar"218$}}
     \raise 2.0pt\hbox{$\mathchar"13C$}}}
\def\gta{\mathrel{\spose{\lower 3pt\hbox{$\mathchar"218$}}
     \raise 2.0pt\hbox{$\mathchar"13E$}}}

\title[Andromeda XXI -- a detailed study]{Andromeda XXI -- a dwarf galaxy in a low density dark matter halo}

\author[M. L. M. Collins et al.]{Michelle L. M. Collins$^{1}$\thanks{E-mail: m.collins@surrey.ac.uk (MLMC)},
Justin I. Read$^{1}$,
Rodrigo A. Ibata$^{2}$,
R. Michael Rich$^{3}$,
\newauthor Nicolas F. Martin$^{2}$, Jorge Pe\~narrubia$^{4}$, Scott C. Chapman$^{5,6}$,  Erik J. Tollerud$^{7}$,
\newauthor Daniel R. Weisz$^{8}$
\\
$^{1}$Physics Department, University of Surrey, Guildford, GU2 7XH, UK\\
$^{2}$Observatoire de Strasbourg,11, rue de l'Universit\'e,
  F-67000, Strasbourg, France\\
$^{3}$Department of Physics and Astronomy, University of
  California, Los Angeles, CA 90095-1547\\
$^{4}$Institute for Astronomy, University of Edinburgh, Royal
  Observatory, Blackford Hill, Edinburgh, EH9 3HJ, UK\\
$^{5}$Eureka Scientific, Inc. 2452 Delmer Street Suite 100, Oakland, CA 94602-3017\\$^{6}$Dalhousie University Dept. of Physics and Atmospheric Science
  Coburg Road Halifax, B3H1A6, Canada\\
$^{7}$ Space Telescope Science Institute, 3700 San Martin Drive, Baltimore, MD 21218, USA
$^{8}$Department of Astronomy, University of California Berkeley, Berkeley, CA 94720, USA\\
}

\date{Accepted XXX. Received YYY; in original form ZZZ}

\pubyear{2020}

\begin{document}
\label{firstpage}
\pagerange{\pageref{firstpage}--\pageref{lastpage}}
\maketitle

\begin{abstract}
Andromeda XXI (And~XXI) has been proposed as a dwarf spheroidal galaxy with a central dark matter density that is lower than expected in the Standard $\Lambda$ Cold Dark Matter ($\Lambda$CDM) cosmology. In this work, we present dynamical observations for 77 member stars in this system, more than doubling previous studies to determine whether this galaxy is truly a low density outlier. We measure a systemic velocity of $v_r=-363.4\pm1.0\kms$ and a velocity dispersion of $\sigma_v=6.1^{+1.0}_{-0.9}\kms$, consistent with previous work and within $1\sigma$ of predictions made using the modified Newtonian dynamics framework. We also measure the metallicity of our member stars from their spectra, finding a mean value of ${\rm [Fe/H]}=-1.7\pm0.1$~dex. We model the dark matter density profile of And~XXI using an improved version of \GravSphere, finding a central density of $\rho_{\rm DM}({\rm 150 pc})=2.6_{-1.5}^{+2.4} \times 10^7 \,{\rm M_\odot\,kpc^{-3}}$ at 68\% confidence, and a density at two half light radii of $\rho_{\rm DM}({\rm 1.75 kpc})=0.9_{-0.2}^{+0.3} \times 10^6 \,{\rm M_\odot\,kpc^{-3}}$ at 68\% confidence. These are both a factor ${\sim}3-5$ lower than the densities expected from abundance matching in $\Lambda$CDM. We show that this cannot be explained by `dark matter heating' since And~XXI had too little star formation to significantly lower its inner dark matter density, while dark matter heating only acts on the profile inside the half light radius. However, And~XXI's low density can be accommodated within $\Lambda$CDM if it experienced extreme tidal stripping (losing $>95\%$ of its mass), or if it inhabits a low concentration halo on a plunging orbit that experienced repeated tidal shocks.
\end{abstract}

\section{Introduction}\label{sec:intro}

In recent years, a number of galaxies with surprisingly low central dark matter densities have been discovered in the Local Group. It has long been observed that low-surface brightness galaxies have lower central densities than predicted by pure cold dark matter (CDM) simulations. Early work showed that their density profiles are more consistent with centrally flat cores rather than the steep cusps predicted by CDM \citep[e.g.][]{flores94,deblok01,read17a}. More recently, comparisons of the dwarf spheroidal (dSph) satellites of  the Milky Way (MW) and Andromeda (M31) with dark matter only simulations show that the observed dSphs also possess a lower density within in their half-light radii ($r_{\rm half}$) as a population than expected, and this is referred to as the ``Too Big to Fail'' (TBTF) problem \citep[e.g.][]{kolchin11,tollerud12,collins14}. For most of these systems, it is likely that repeated gravitational potential fluctuations, driven by feedback from star formation, has played a role in softening their central cusps over time into flatter cores \citep[e.g.][]{navarro96,read05,pontzen12,zolotov12,brooks14,onorbe15,read16,read19a}. This process -- that has become known as `dark matter heating' -- is only effective in galaxies above some stellar mass-to halo mass ratio threshold ($M_*/M_{200} \sim 5 \times 10^{-4}$; \citealt{dicintio14a,dicintio14b}), and for systems with extended star formation \citep{read16,read19a}.

However, the low density problem seems particularly acute in some of the more diffuse companions of the Milky Way and Andromeda. Around our Galaxy, two extremely diffuse satellites -- Antlia 2 and Crater II \citep{torrealba16a,torrealba19a} -- have been uncovered, whose half light radii far exceed that expected from their similarly bright counterparts. Both objects have a surface brightness of $\mu_0>30$~mag~arcsec$^{-2}$, and effective radii in excess 1~kpc ($r_{\rm half}= 1.1$ and $2.9\,{\rm kpc}$ respectively). Similarly, in M31, the extreme object Andromeda XIX was discovered by \citet{mcconnachie08} in the Pan-Andromeda Archaeological Survey, with $\mu_0=29.2\pm0.4$~mag~arcsec$^{-2}$ and $r_{\rm half}=3.1^{+0.9}_{-1.1}\,{\rm kpc}$ \citep{martin16c}. All three of these ``feeble giants'' also appear to reside in surprisingly low mass halos. Through measured velocity dispersions, their mass within $r_{\rm half}$ are far lower than expected for systems of their size or brightness \citep{caldwell17,fu19,torrealba19a,collins20}, raising questions about whether they can be understood in the context of the $\Lambda$ Cold Dark Matter ($\Lambda$CDM) framework. For these systems, it is unlikely that star formation feedback alone can explain their low densities. Thus far, detailed star formation histories are not available for these objects, but even assuming highly efficient dark matter heating, their current sizes, surface brightness and velocity dispersions cannot be reproduced \citep{torrealba19a}. Instead, it is assumed they must also have undergone extreme tidal interactions while orbiting their host galaxy (e.g. \citealt{fattahi18}). 

There are two main tidal effects that can act to lower the density of orbiting satellites: tidal stripping and tidal shocking. Tidal stripping occurs when the tidal force from the host galaxy exceeds that from the satellite, causing dark matter and/or stars to become unbound \citep[e.g.][]{vonhoerner57,read06a}. Since this peels away stars and dark matter from the outside in, it only lowers the inner density after extreme stripping has occurred. For cuspy dark matter profiles, this means losing >99\% of the satellite's initial mass \citep{penarrubia08b,penarrubia10,errani20}. Shallower or cored dark matter profiles require less extreme mass loss \citep{read06b,penarrubia10,brooks14}. Tidal shocking occurs for satellites moving on eccentric orbits if the external gravitational field changes more rapidly than the internal dynamical time of the stars and/or dark matter. This means that tidal shocks are maximised at pericentre, where the external field changes most rapidly \citep[e.g.][]{spitzer58,gnedin99}, and for low density satellites, since the internal orbit time goes as $\rho^{-1/2}$, where $\rho$ is the total density. For this reason, in the context of $\Lambda$CDM, satellites on plunging orbits can only efficiently lower their density through tidal shocks if they have a central dark matter core \citep[e.g.][]{read06b,errani17,errani20,vdb18}, or if they inhabit low-concentration dark matter halos before infall \citep{amorisco19a}. Unlike dark matter heating, the combination of tidal stripping and shocking will lower the satellite's density at all radii, not just in the centre \citep[e.g.][]{kazantzidis04b,read06b}. However, it can be challenging to detect, observationally, when this has occurred. The tell-tale signatures of tides -- stellar streams, distorted outer stellar isophotes and/or tangential anisotropy -- all manifest in the low surface brightness stellar outskirts \citep[e.g.][]{read06b,ural15,amorisco19a,genina21}.

For Antlia 2 and Crater II, proper motions can be measured from Gaia DR2 \citep{fritz18,fu19,torrealba19a}, and both are consistent with being on radial orbits that bring them within a few 10s of kpc from the Galactic centre -- the regime in which tidal shocks are likely important \citep{read06b}. As such, it has been argued that tidal processes govern the evolution of these galaxies \citep{sanders18,amorisco19a}. Indeed, a plunging orbit for Antlia 2 could also explain observed disturbances in the outer HI disc of the Milky Way \citep{chakrabarti19}. A precise orbit for Andromeda~XIX has not been determined, but from its structural and dynamical properties, it is possible that this galaxy is also on a similarly plunging orbit \citep{mcconnachie08,martin16c,collins13,collins20}. To determine whether {\it all} low mass extended galaxies can be explained in this manner in $\Lambda$CDM, however, we need to study more systems.

In this work, we turn our attention to another significant low-mass outlier, Andromeda XXI (And~XXI). This M31 satellite has a luminosity of $L=3.2\times10^5\,{\rm L_\odot}$ and a half-light radius of $r_{\rm half}=1033^{+206}_{-181}\,{\rm pc}$ \citep{martin16c,weisz19b}, approximately 3 times more extended than other galaxies of comparable luminosity. And, based on an earlier dynamical study, it has a low central velocity dispersion of $\sigma_v=4.5^{+1.2}_{-1.0}\,\kms$, consistent with residing in a very low density halo \citep{collins13,collins14}. However, these findings were based on dynamics of only 32 member stars, meaning that detailed dynamical modelling of its halo could not be carried out. In this work, we reanalyse the mass for And~XXI using dynamics for 77 member stars, modelling its radial dark matter density profile for the first time with an updated version of the \GravSphere\ Jeans code \citep{read17a,read17b,read18a,gregory19a,genina20}. We combine our dynamical analysis with the measured star formation history for this galaxy from \citet{weisz19a} to determine whether its central density can be explained by dark matter heating driven by stellar feedback \citep[e.g.][]{navarro96b,read05,pontzen12,read16,read19a}, and to assess what role -- if any -- tidal interactions may play in explaining its properties.

This paper is organised as follows. In \S2, we start by discussing our observations. We then present the observed dynamics and metallicity of our member stars in \S3. In \S4, we describe several improvements we have made to the \GravSphere\ code to improve its performance for modelling systems with small numbers of stars, and we present our dynamical modelling of And~XXI (tests of our new method on mock data are included for completeness in Appendix \ref{app:mocks}). In \S5, we discuss our results and set them in the context of prior work in the literature. Finally, in \S6 we present our conclusions.

\section{Observations}

\subsection{DEIMOS Spectroscopy}

\begin{table*}
	\centering
	\caption{Details of And~XXI spectroscopic observations (PI Rich). A total of 88 And~XXI velocities were measured, for 77 independent stars (11 repeat measurements).}
	\label{tab:specobs}
	\begin{tabular}{lcccccccc}
		\hline
		Mask name & Date & RA & Dec  & Position angle (deg)& Exposure time ($s$) & No. targets & No. members \\
		\hline
7And21 & 26 Sep 2011 & 23:54:47.70  & 42:28:33.6 & 180 & 3600 & 157 & 32 \\
A21maj & 01 Oct 2013 & 23:54:47.70   & 42:28:15.0 & 147 & 7200 & 112 & 26\\
A21min & 01 Oct 2013 & 23:54:47.70  & 42:28:15.0 & 57 & 7200 & 110 & 30\\
		\hline
	\end{tabular}
\end{table*}

Spectroscopic observations of And XXI were undertaken using the
Deep Extragalactic Imaging Multi-Object Spectrograph (DEIMOS, \citealt{deep2,cooper12}), which is mounted on the
Keck II telescope in Mauna Kea. The multi-object mode
of DEIMOS allows us to simultaneously observe $\sim150$ stellar targets in a
single pointing, spread across the significant field of view of DEIMOS
($16^{\prime}\times8^{\prime}$, which translates to $\sim3.5\kpc\times1.3$~kpc
at the distance of Andromeda). These aspects make DEIMOS ideal for surveying
And XXI, whose half-light radius of $1033^{+206}_{-181}\,{\rm pc}$ \citep{martin16c,weisz19b} fits comfortably within a single mask. The observations presented in this
work are taken from 3 DEIMOS masks, which were observed in September 2011
(previously presented in \citealt{collins13}) and October 2013 as part of the
Z-PAndAS spectroscopic survey (PI Rich). The instrumental set-up for each mask was
identical, and used the 1200 line mm$^{-1}$ grating (resolution of 1.4\AA\
FWHM). A central wavelength of 7800\AA\ was used, allowing us spectral coverage
from $\sim5600-9800$\AA. This permitted us to resolve the region of the
calcium triplet (Ca {\sc II}) at $\lambda\sim8500$\AA, a strong absorption
feature that we use to determine both velocities and metallicities for our
observed stars. The exposure time for the mask observed in 2011 was 3600s (in
$3\times1200$s sub-exposures), while the two 2013 masks were observed for
7200s (in $6\times1200$s sub-exposures). The average seeing per mask was
0.6$^{\prime\prime}$, 0.8$^{\prime\prime}$ and 0.7$^{\prime\prime}$,
resulting in typical S:N values of $\sim5$ per pixel. We present the details of each observed mask in table~\ref{tab:specobs}.

We reduce the resulting science spectra using a custom built pipeline, as
described in \citet{ibata11} and \citet{collins13}. Briefly, the pipeline
identifies and removes cosmic rays, corrects for scattered light, illumination,
slit function and fringing before performing flat-fielding to correct for
pixel-to-pixel variations.  We perform a wavelength calibration of each pixel
using arc-lamp exposures. Finally, we subtract the sky from the 2 dimensional
spectra before extracting each spectrum -- without resampling -- in a small
spatial region around the target. We then derive velocities for all our stars
using the Ca {\sc II} triplet absorption feature.  The velocities are
determined using a Markov Chain Monte Carlo procedure where a template Ca {\sc
  II} spectrum was cross-correlated with the non-resampled data, generating a
most likely velocity for each star, and a likely uncertainty based on the
posterior distribution that incorporates all the uncertainties for each
pixel. Typically our velocity uncertainties lie in the range of 3-15
kms$^{-1}$. Finally, we also correct these velocities to the heliocentric
frame.

As we aim to combine kinematic data from three masks, observed at different
times and in different conditions, it is imperative that we correct our final
velocities for any systematic shifts caused by mis-alignments of stars within
the slits themselves. Such misalignments can be caused by astrometry errors or
from a slight offset of the position angle of the mask on the sky, and can
introduce velocity shifts in our spectra of up to $\sim15\kms$
\citep{collins13}. Such shifts can be corrected for using atmospheric telluric
absorption lines, which are imprinted on each of our spectra. As they
originate from the atmosphere, they should always be observed at their
rest-frame wavelength. By cross-correlating the science spectra with a model
telluric spectrum, one can measure and correct for any velocity shifts
introduced by the types of mis-alignments mentioned above. As a further check for any mis-alignments, we also look for signatures of velocity gradients along each of our masks for our likely And~XXI member stars. We find no evidence of any gradients. We also check for any change in velocity uncertainty with mask position, and again find no correlation.

Within our catalogue, we have 8 stars with more than one velocity measurement. We use these to check our velocity calibrations described above are reasonable. We find a mean (median) offset velocity for these stars of $v_{\rm offset}=1.2~(2.8)\,\kms$, both of which are within our typical velocity uncertainties. As we have only a small handful of duplicates, we cannot use these to re-derive an independent measurement of the systemic velocity uncertainty inherent to DEIMOS. Instead, we use the value determined from observations of stars in Galactic satellites by \citet{simon07} of $2.2\,{\rm km\,s^{-1}}$. We find the velocity differences for our 8 duplicates to be fully consistent with this value. We combine this in quadrature with the individual velocity uncertainties described above. We could instead use the value of $3.4\,{\rm km\,s^{-1}}$ from \citet{martin14b}, which was derived from observations of M31 dSphs reduced with the same pipeline as this work. However, using a higher value will bias us to a slightly lower value for the dispersion of And XXI. As we are aiming to confirm whether this system is low mass, we find the \citet{simon07} value to be a more conservative approach. We provide all measured velocities, metallicities and other reduced properties for our member and non-member stars in an electronic file, available online at the journal website. The reduced spectra are also available upon request, while the raw data are available through the Keck archive.

\subsection{Subaru SuprimeCam imaging}

Subaru Suprime-Cam imaging of And~XXI were undertaken on 21-22 August 2009 in Cousins $V$ and $i_c$ filters, in photometric conditions with an average seeing of $\sim1^{\prime\prime}$. A single field was observed, with $5\times440\,s$ in the $V-$band, and $20\times240\,s$ in the $i_c-$band. The data were processed with the CASU pipeline \citep{irwin01}, which debiased, flat-fielded, trimmed and gain corrected the images. A catalogue was generated, and each source was morphologically classified as either stellar, non-stellar or noise-like. Finally, as we wished to combine these with PAndAS imaging of And~XXI \citep{martin16c}, we transformed these into CFHT-MegaCam $g$ and $i-$band magnitudes, which also allowed for a full calibration of our catalogue, using the following calibration,

\begin{eqnarray}
g &=& V - 0.044\times(V-i_c)+0.060 \\
i& =& i_c + 0.033\times(V-i_c)+0.036. 
\end{eqnarray}

\begin{figure}
  \begin{center}
     \includegraphics[angle=0,width=0.95\hsize]{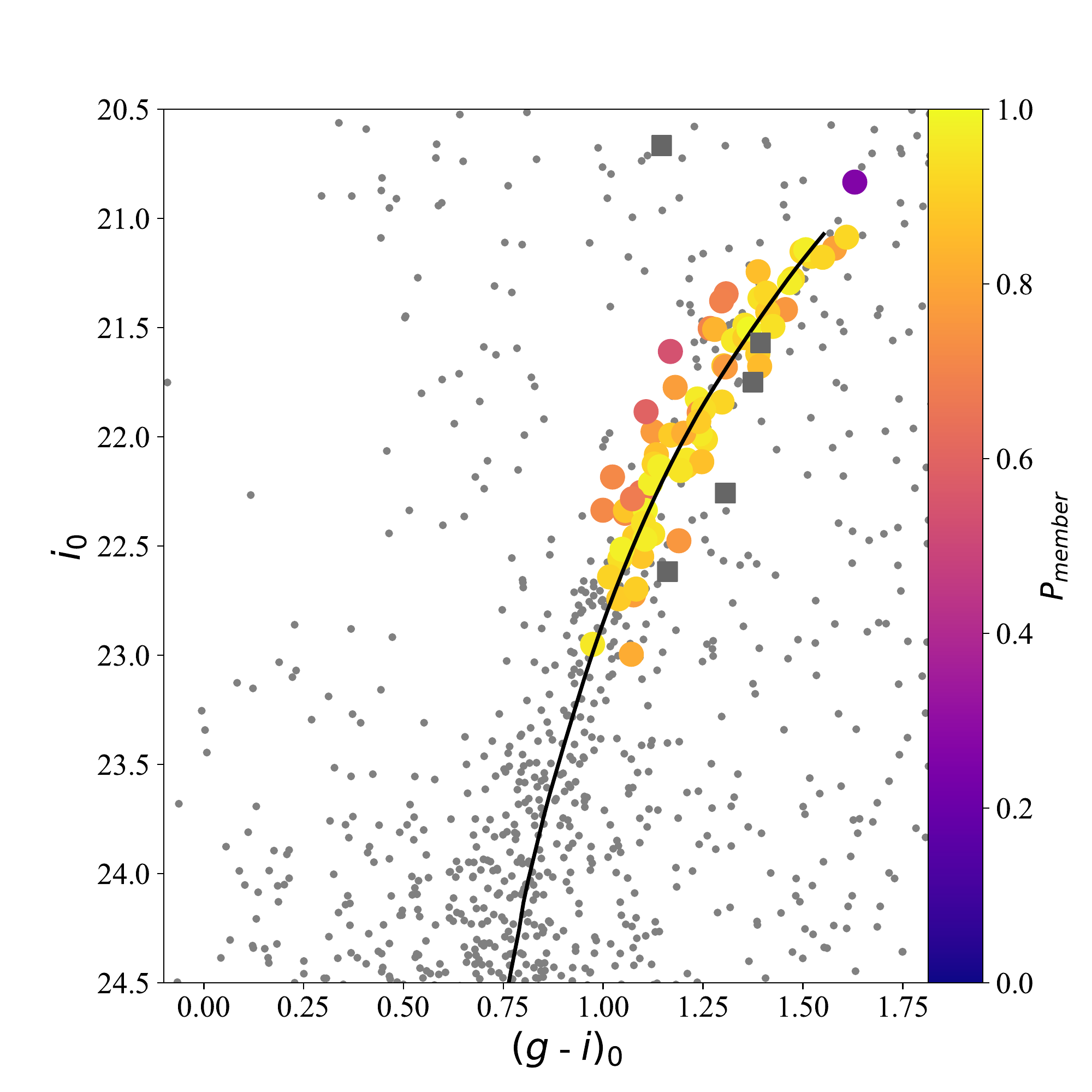}
 \caption{A colour magnitude diagram for And ~XXI, constructed from Subaru Suprime-cam data. Likely members of And~XXI with spectroscopic data are highlighted as large points, colour-coded by probability of membership (derived from imaging and spectroscopic data). The red giant branch can be clearly seen. The solid line is an isochrone from the Dartmouth database with [Fe/H]$=-1.8$, $[\alpha/$Fe]$=0.4$, age = 12~Gyr, shifted to the distance modulus of And ~XXI, $m-M=24.6$ \citep{dotter08,conn13}.  The stars in the possible M31 stream are highlighted as grey squares. 4 of them lie along a similar RGB which would be well described with an isochrone of [Fe/H]$\sim-1.5$.}
\label{fig:cmd} 
  \end{center}
\end{figure}

Finally, we extinction correct all our data using the dust maps of \citet{schlegel98}. Our final colour magnitude data for these data are shown in fig.~\ref{fig:cmd}.

\section{Results and analysis}

In this section we present our measurements of the basic chemodynamic properties of And~XXI. All numerical results are reported in table~\ref{tab:a21prop}. Throughout our analysis we make use of various python packages, specifically {\sc NumPy, SciPy, Astropy} and {\sc Matplotlib} \citep{numpy,scipy, astropy,matplotlib}.

\begin{table}
	\centering
	\caption{The properties of And~XXI.}
	\label{tab:a21prop}
	\begin{tabular}{ll} 
		\hline
		Property &  \\
		\hline
		$\alpha,\delta$ (J2000)$^1$ & 23:54:47.9, +42:28:14 \\
		$m_{V,0}$$^a$ & $15.5\pm0.3$ \\
		$M_{V, 0}$ $^a$& $-9.1\pm0.3$ \\
		Distance (kpc) $^b$& $827^{+23}_{-25}$\\
		$r_{\rm half}$ (arcmin) $^a$& $4.1^{+0.8}_{-0.4}$ \\
		$r_{\rm half}$ (pc)$^a$ & $1005\pm175$\\
		$\mu_0$ (mag per sq. arcsec)$^a$ & $28.0\pm0.3$\\
		$L$ ($L_\odot$)$^a$ & $3.2^{+0.8}_{-0.7}\times10^5$ \\
		$v_r (\kms)$$^c$ & $-363.4\pm1.0$~kms$^{-1}$\\
		$\sigma_v (\kms)$ $^c$&  $6.1^{+1.0}_{-0.9}$~kms$^{-1}$\\
		$M(r<1.8r_{\rm half} (M_\odot)^c $& $3.0^{+0.9}_{-0.8}\times10^7$\\
		$[M/L]_{\rm half} (M_\odot/L_\odot)^c$ & $78^{+30}_{-28}$ \\
		$[{\rm Fe/H}]$ (dex) $^c$& $-1.7\pm0.1$ \\
		\hline
	\end{tabular}
	\\$^a$ \citet{martin16c},
	$^b$ \citet{conn12b},
	$^c$ This work
\end{table}

\subsection{Classifying And XXI member stars}
\label{sect:mem}

To investigate the mass profile of And XXI, we first identify which stars observed with DEIMOS are probable members following the procedure outlined in \citet{collins13} and \citet{collins20}. Briefly, this method assigns probability of membership based on three criteria: (1) the stars
position on the colour magnitude diagram of the dwarf galaxy, $P_{\mathrm {CMD}}$ (2) the distance of the star from the
centre of the dwarf galaxy $P_{\rm dist}$ and (3) the velocity of the star,
$P_{\rm vel}$. The probability of membership can then be expressed as a
multiplication of these three criteria:

\begin{equation}
P_{\mathrm {member}}\propto P_{\mathrm {CMD}}\times P_{\mathrm{ dist}}\times P_{\mathrm {vel}}
\end{equation}

$P_{\rm
  CMD}$ is determined using the colour magnitude diagram (CMD) of And XXI. We
implement a method based on that of \citet{tollerud12}, using an isochrone to isolate those stars most likely to be associated with And~XXI (see figure~\ref{fig:cmd}). We use an old, metal poor isochrone from the {\sc Dartmouth} stellar evolutionary models ([Fe/H]$=-1.8$, $[\alpha/$Fe]$=0.4$, age = 12~Gyr, shifted to a distance modulus of $m-M=24.65$ \citealt{dotter08,weisz19b}) that well represents the RGB of the dwarf galaxy. We then measure the minimum distance of a star from this isochrone ($d_{\rm min}$), and assign a probability using the following equation:

\begin{equation}
P_{\rm CMD}=\exp \left(\frac{- d_{\rm min}^2} {2  \sigma_{\rm CMD}^2}\right)
\label{eq:pcmd}
\end{equation}

\noindent where $\sigma_{\rm CMD}=0.05$. 
$P_{\rm dist}$ is determined using
the known radial surface brightness profile of the dwarf, modelled as a
Plummer profile, using the half-light radius and ellipticity parameters for
And XXI as determined from PAndAS data \citep{martin16c}. $P_{\rm vel}$
is determined by simultaneously fitting the
velocities of all observed stars assuming that 3 dynamically distinct, Gaussian components are present: the MW
foreground contamination ($P_{\rm MW}$, with systemic velocity $v_{\rm MW}$ and velocity dispersion of $\sigma_{v,\mathrm{MW}}$), the M31 halo
contamination ($P_{\rm M31}$, with systemic velocity $v_{\rm M31}$ and velocity dispersion of $\sigma_{v,\mathrm{M31}}$), and
And~XXI, $P_{\rm A21}$, with an arbitrary systemic velocity, $v_r$ and velocity dispersion
$\sigma_v$. As such, the probability that a given star belongs to component $y$ (where $y$ is the MW, M31 or And XXI) is

\begin{equation} 
\begin{aligned}
  P_{y,i}=\frac{1}{\sqrt{2\pi(\sigma_{v,y}^2+\delta_{vr,i}^2)}}\times{\rm exp}\left[-\frac{1}{2}\left(\frac{v_y-v_{r,i}}{\sqrt{\sigma_{v,y}^2+\delta_{vr,i}^2}}\right)^2\right] \end{aligned},
\end{equation}

\noindent where $v_{r,i}$ and $\delta_{vr,i}$ are the velocity and uncertainty for the $i-$th star. In reality, the Milky Way distribution is more complex than a single Gaussian assumes (see, e.g. \citealt{gilbert06,collins13}), however as And XXI is well-separated kinematically  from the Milky Way population, this assumption does not affect our membership determination. The overall likelihood function can then be simply written as a combination of these three components,

\begin{multline}
\mathcal{L}_i(v_{r,i}, \delta_{vr,i}|\mathcal{P}) = (1 - \eta_{\rm MW} - \eta_{\rm M31}) \times P_{\rm A21}\\
+ \eta_{\rm MW} \times P_{\rm MW} + \eta_{\rm M31} \times P_{\rm M31},
\label{eqn:pdfsimple}
\end{multline}

\noindent where $\eta_{\rm MW}$ and $\eta_{\rm M31}$ are the fraction of our sample found within the Milky Way and M31 halo components of the model. We use the MCMC \texttt{emcee} code \citep{fm13a} to explore a broad parameter space for these components. We set uniform priors for each of our parameters (the velocities and dispersions for each population, see table~\ref{tab:priors} for details). In addition,  we set $0<\eta_{\rm MW/M31}<1$. The dynamical values we measure using this technique will likely resemble the final values derived from a probability weighted analysis, but without including prior information about CMD and spatial positions of stars, these may be biased by the inclusion of contaminant M31 halo stars. This is thus the first step in determining the kinematic parameters. This first-pass kinematic analysis gives $v_r-363.6\pm1.0\kms$ and $\sigma_v=6.2^{+1.0}_{-0.9}\kms$ for And~XXI.

\begin{table}
	\centering
	\caption{Prior values used in our {\sc emcee} analysis}
	\label{tab:priors}
	\begin{tabular}{lll} 
			\hline
		Component & Prior &  \\
		   & $v_{r}\, (\kms)$ prior  & $\sigma_v \,(\kms)$  \\
		\hline
		And~XXI & $-400<v_{r}<-340$ & $0<\sigma_{v}<50$ \\
		Milky Way & $-90< v_{r}<0$ & $0<\sigma_{v}<150$\\
		M31 & $-340<v_{r}<-220$ & $0<\sigma_{v}<150$ \\
		\hline
	\end{tabular}
\end{table}
 
\begin{figure*}
  \begin{center}
    \includegraphics[angle=0,width=0.45\hsize]{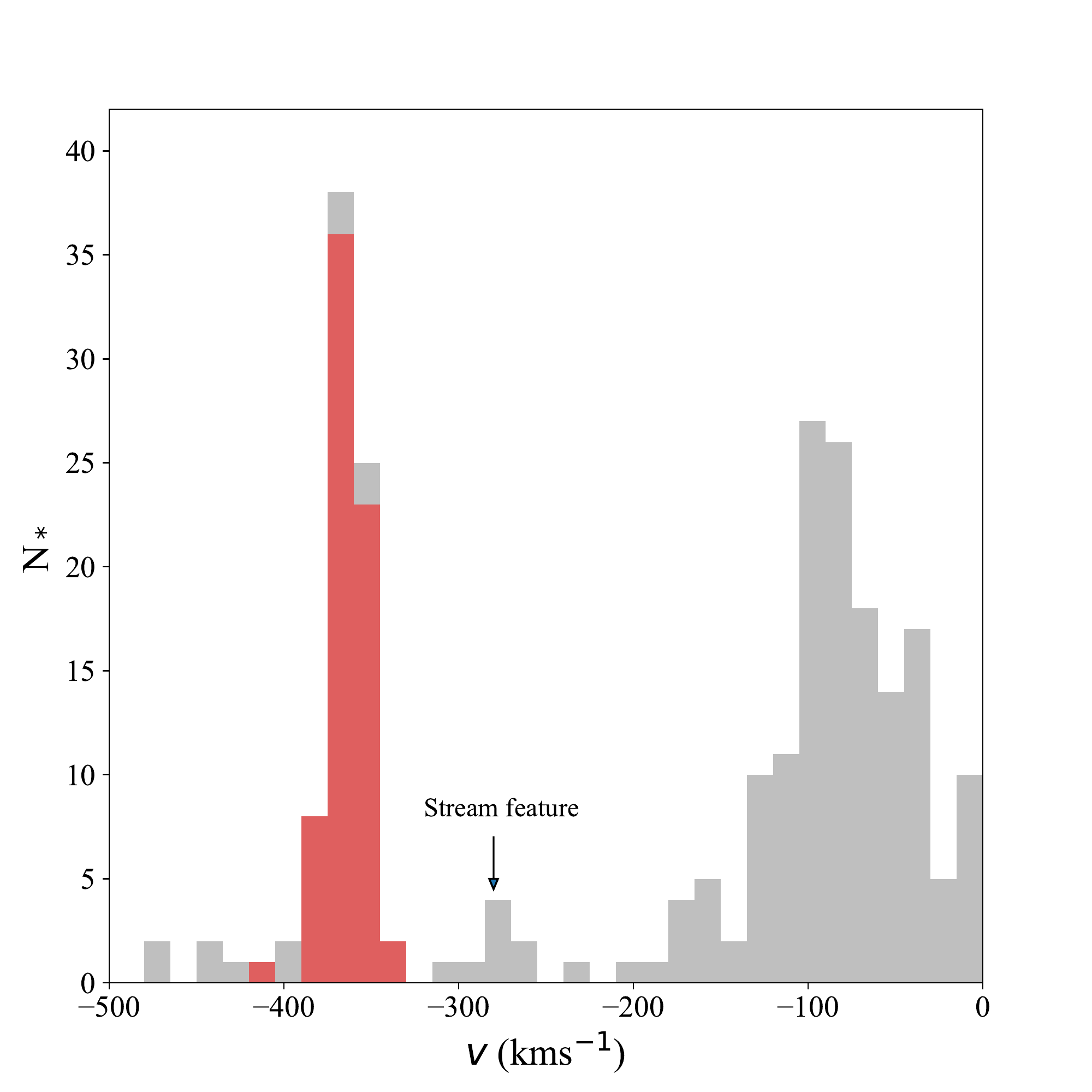}
\includegraphics[angle=0,width=0.45\hsize]{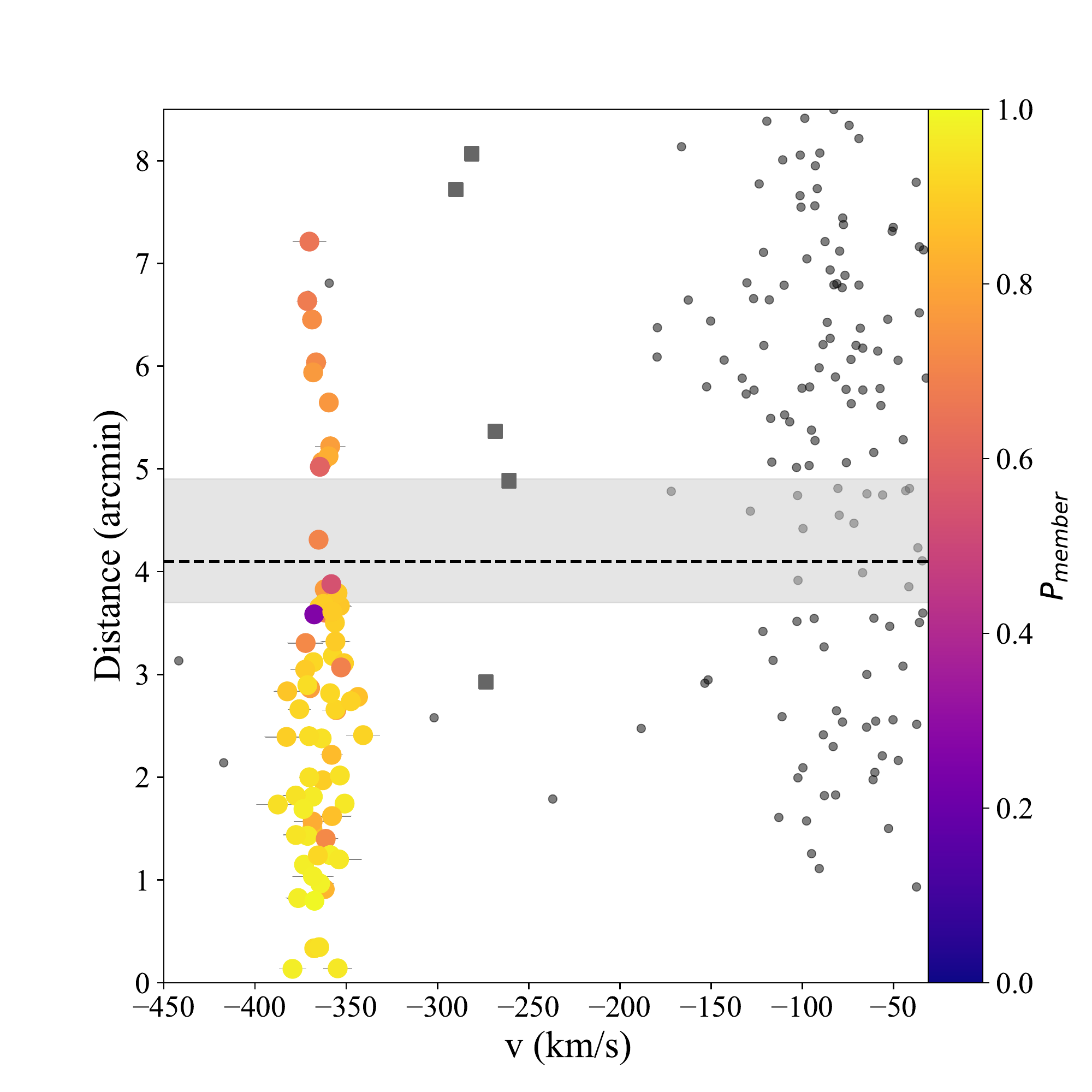}
\caption{{\bf  Left:} Velocity histogram of all stars observed in our 3 DEIMOS
  fields, highlighting And XXI members ($P_{\rm member}>0.1$) in red. A kinemtically cold feature in the M31 halo is also highlighted.{\bf Right:} The distance from the centre of And XXI as a
  function of velocity. All probable And~XXI members are colour coded by probability of membership. The half-light radius is indicated as a dashed line, with the grey shaded region indicating the $1\sigma$ uncertainties on this value \citep{martin16c}. Interestingly, a clear difference can already be seen at this boundary, where the stars within $r_{\rm half}$ show a significantly larger dispersion than those beyond $r_{\rm half}$. The stars in the possible M31 stream are highlighted as grey squares.}
  \label{fig:vels}
  \end{center}
\end{figure*}

We then combine these three probabilities ($P_{\rm CMD}, P_{\rm dist}, P_{\rm A21}$) and calculate the
likelihood of a given star being a member of And XXI. The results of this analysis are shown in fig.~\ref{fig:vels}, where we display the
kinematic distribution of all stars in the left hand panel, with the most
likely stars highlighted in red. The right panel shows the distance of the
stars from the centre of And XXI as a function of their velocity, colour coded by their probability of membership, $P_{\rm
  member}$. The dashed line represents $r_{\rm half}$. Already an interesting trend can be seen here, as the stars within the half-light radius show a far broader dispersion than those located further out. While only 10 stars contribute to this outer population, it is certainly striking.
  
In total, we identify 77 stars with $P_{\rm member}>0.1$ as probable members of And XXI (a factor 2.5 increase compared with 32 in our previous study, \citealt{collins13}). All our non-members have $P_{\rm member}<<0.1$, significantly lower than our lowest probability member. We find that our results are insensitive to this choice, as all but 5 members have $P_{\rm member}>0.6$.

\begin{figure}
  \begin{center}
     \includegraphics[angle=0,width=0.95\hsize]{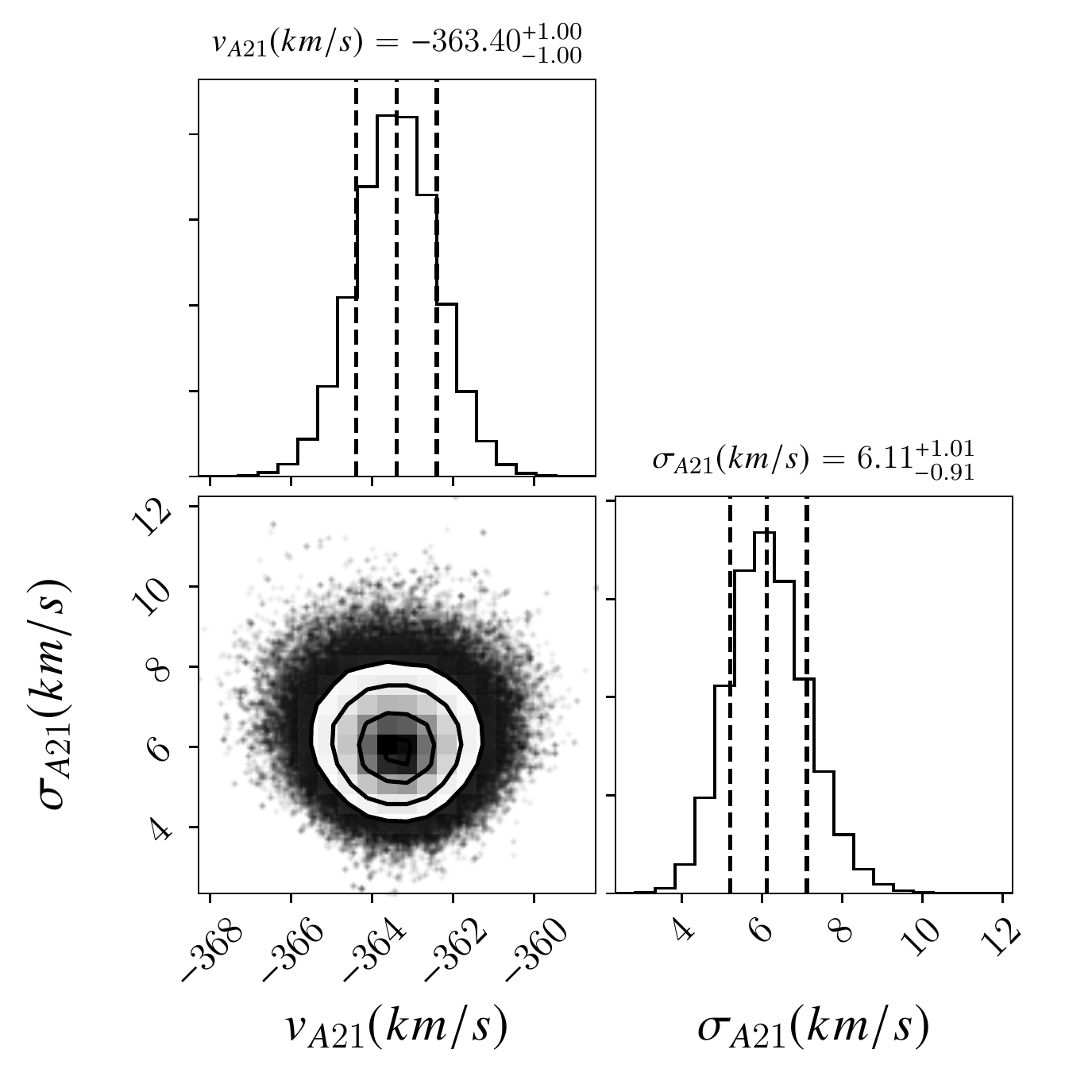}
  \caption{A corner plot showing the results of our kinematic analysis using {\sc emcee}. Both velocity and velocity dispersion are well resolved, giving $v_r=-363.4\pm1.0$ and $\sigma_v=6.1^{+1.0}_{-0.9}$.}
  \label{fig:MCMC}
  \end{center}
\end{figure}

We next use these probabilities as weights in our analysis. To determine the systemic velocity ($v_r$) and velocity dispersion ($\sigma_v$) of And~XXI, we define a likelihood function, $\mathcal{L}$, that describes a single Gaussian population: 

\begin{multline}
\log~\mathcal{L}_{A21}\left(v_{r,i}|\langle v_r \rangle,\theta,\sqrt{\sigma_{vr}^2 + \delta_{vr,i}^2}\right) = -\frac{1}{2}\sum_{i=0}^N  \log(\sigma^2) \\
+ \left(\frac{(v_r - v_{r,i})^2}{2\sigma^2}\right) +  \log(2\pi) + \log(P_{{\rm member}, i})
\label{eqn:grad}
\end{multline}

\noindent where $\sigma=\sqrt{\sigma_{v}^2+\delta_{v,i}^2}$ is the combination of the underlying velocity dispersion of And~XXI and the velocity uncertainty of individual stars ($\delta_{v,i}$). $P_{{\rm member}, i}$ is the probability of membership of the $i-$th star. We again use {\tt emcee} to investigate the parameter space, using the results from our first pass as the initial starting guess for And~XXI's parameters, and the same uniform priors. The resulting posterior distribution can be seen in fig.~\ref{fig:MCMC}. Both the systemic velocity and velocity dispersion are well resolved, giving median values of  $v_r=-363.4\pm1.0\kms$ and $\sigma_v=6.1^{+1.0}_{-0.9}\kms$, where the uncertainties are the 68\% percentiles of the posterior distributions. These values are nearly identical to our 3 component, kinematic-only approach above. They are also consistent with C13 values derived from 32 stars of  $v_{r,C13}=-362.5\pm0.9\kms$ and $\sigma_{v,C13}=4.5^{+1.2}_{-1.0}\kms$. Given this similarity, it is likely that And~XXI still resides in a low mass halo, as found by C13. We return to this issue in \S~\ref{sec:mass}.

To investigate the visually striking change in dispersion at $r_{\rm half}$ seen in fig.~\ref{fig:vels}, we run the same MCMC process on stars inside and outside this boundary. The results for $v_r$ are fully consistent for both samples, but $\sigma_v$ changes dramatically. For the 67 central stars, we measure $\sigma_v(r<r_{\rm half})=6.9^{+1.1}_{-1.0}\,\kms$, consistent with our main analysis. However, for the 10 stars at larger radii we find $\sigma_v(r>r_{\rm half})=1.8^{+2.0}_{-1.2}\,\kms$. This is a puzzling difference. As we only have 10 stars in the outer sample, it is hard to draw a strong conclusion about this significant drop at this time.

\subsection{The kinematic distribution of And XXI}
\label{sect:kin}

\begin{figure}
  \begin{center}
     \includegraphics[angle=0,width=0.9\hsize]{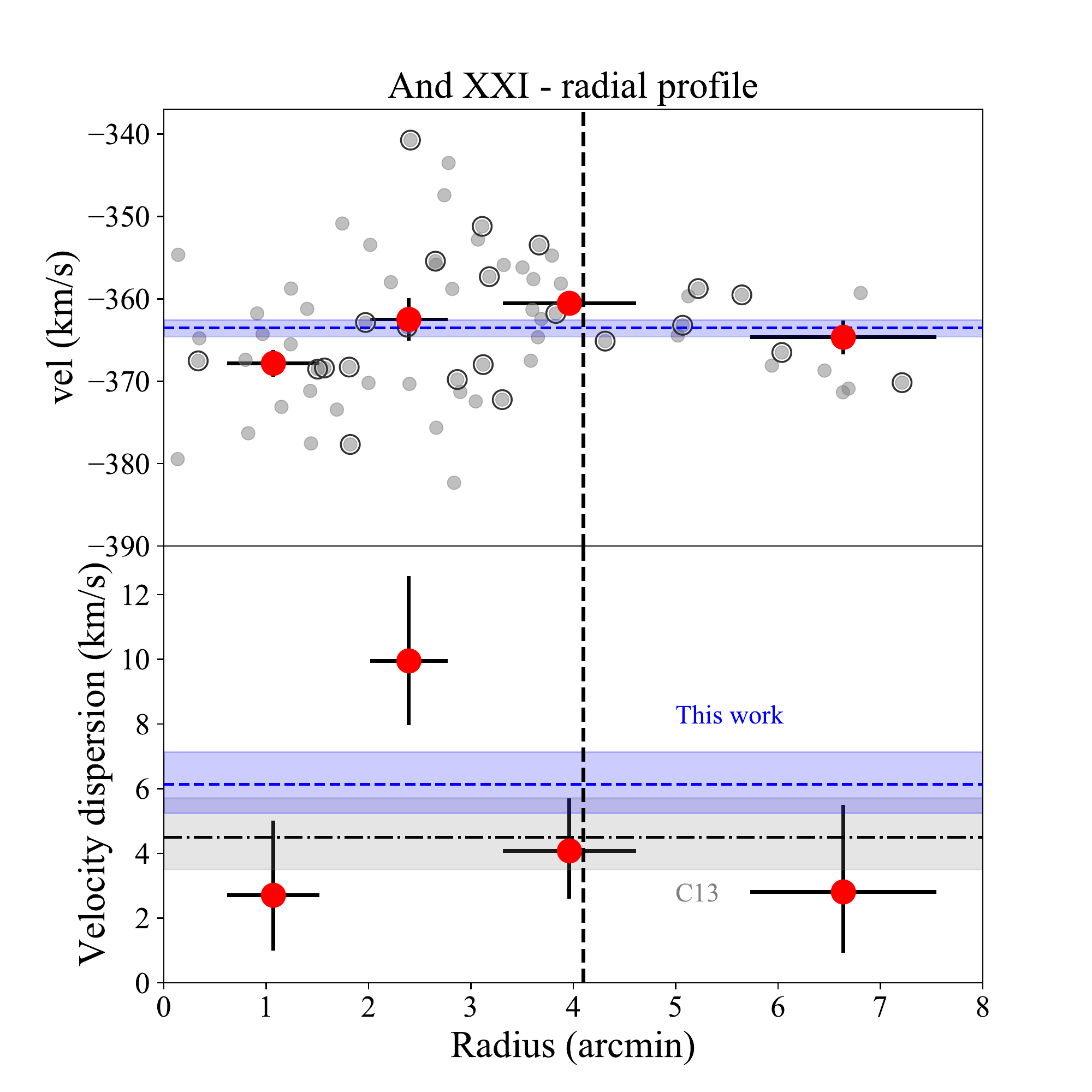}
     \caption{Radial velocity profile (top) and velocity
       dispersion profile (bottom) for And XXI. Individual stellar velocities are also shown in the top panel as small grey points, with those from our original survey highlighted as encircled points. The blue dashed lines and shaded regions represent the
       systemic velocity and dispersion for the whole dataset, plus their $1\sigma$ confidence intervals as calculated
       using our MCMC approach. In the lower panel, we also show the average dispersion derived from our previous work (C13) as a dot-dashed line and grey shaded region. Our new value is slightly higher, but perfectly consistent within the uncertainties with C13. From the velocity profile, it
       appears that this system is not in equilibrium, as variations around
       the systemic are often significant ($>1-2\sigma$). The dispersion
       profile is also surprising. As all our data are within $\sim2\times
       r_{\rm half}$, one expects the profile to remain constant as a function
       of radius owing to the extended dark matter halo. Instead, we see that the majority of the profile is flat, with a dispersion of $\sim3-5\kms$, with the exception of a far hotter bin at $\sim2^{\prime}$. This bin has a dispersion that is inconsistent with the others at a level of $\sim2\sigma$.}
  \label{fig:radial}
  \end{center}
\end{figure}

With our sample of 77 stars in And XXI, we can measure how the radial velocity and velocity dispersion
behave as a function of radius, and attempt to map its mass profile. Typically, dSphs are dispersion supported, with little to no rotation, and we expect them to have a reasonably constant systemic velocity (and velocity dispersion) as a function of radius. In
fig.~\ref{fig:radial} we show how the systemic velocity (top) and velocity
dispersion (bottom) vary as a function of radius in the system. We construct this
figure by arranging our stars into 5 equal-sized bins, each with 19
stars. We only include stars with $P_{\rm member}>0.1$. In each bin, we compute $v_r$ and $\sigma_v$ using the same technique
 as before (equation~\ref{eqn:pdfsimple}). What is immediately striking is that
there is a non-symmetric variation of both these quantities with
respect to the global values (dashed lines) derived in \S~\ref{sect:mem}. From
the velocity profile, we note that the $v_r$ determined for the first and third bins (at
$\sim1$ and $3.5$~arcmin) are inconsistent with one another, and the systemic
velocity at a significance of $>2\sigma$. Such fluctuations may indicate that
this is not a system in dynamical equilibrium.

Equally striking is the behaviour of the velocity dispersion as a function of
radius. As dSph galaxies are thought to be deeply embedded within extended
dark matter halos, one expects the velocity dispersion of the galaxy to trace
this component. As such, it should remain constant as a function of radius out
to many half-light radii. However, we have already seen that the dispersion in the outskirts of And~XXI is significantly lower than the central value, and this is further demonstrated here. What we see is that the majority of the stellar population has a dispersion consistent with $\sim2-4\kms$, with the exception of one bin at $\sim2^\prime$ ($\sim480$~pc), which appears significantly dynamically hotter. This can even be seen in the raw data shown in fig.~\ref{fig:vels}. Is this some signature of substructure, or merely an artefact of small number statistics? 

\begin{figure*}
  \begin{center}
     \includegraphics[angle=0,width=0.45\hsize]{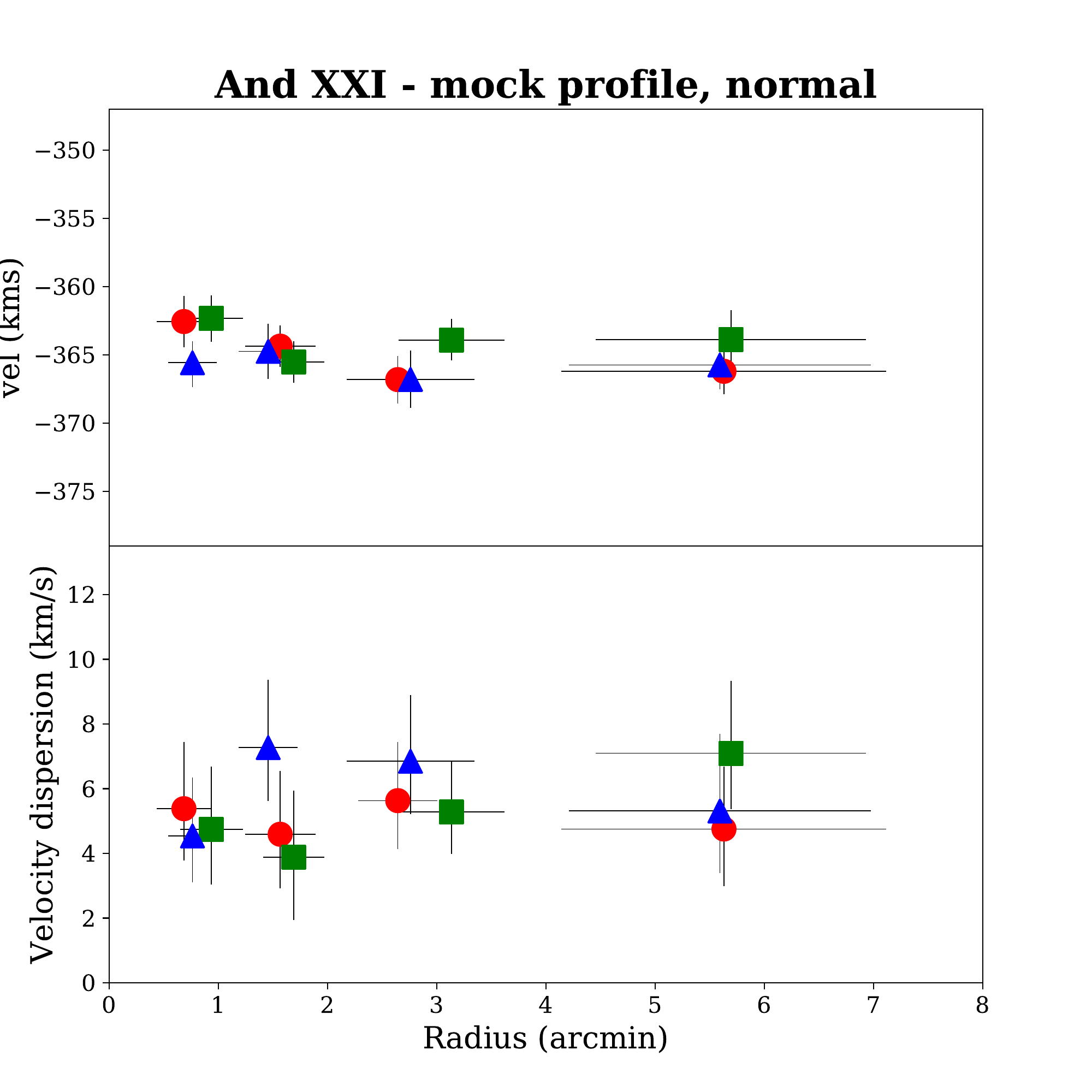}
     \includegraphics[angle=0,width=0.45\hsize]{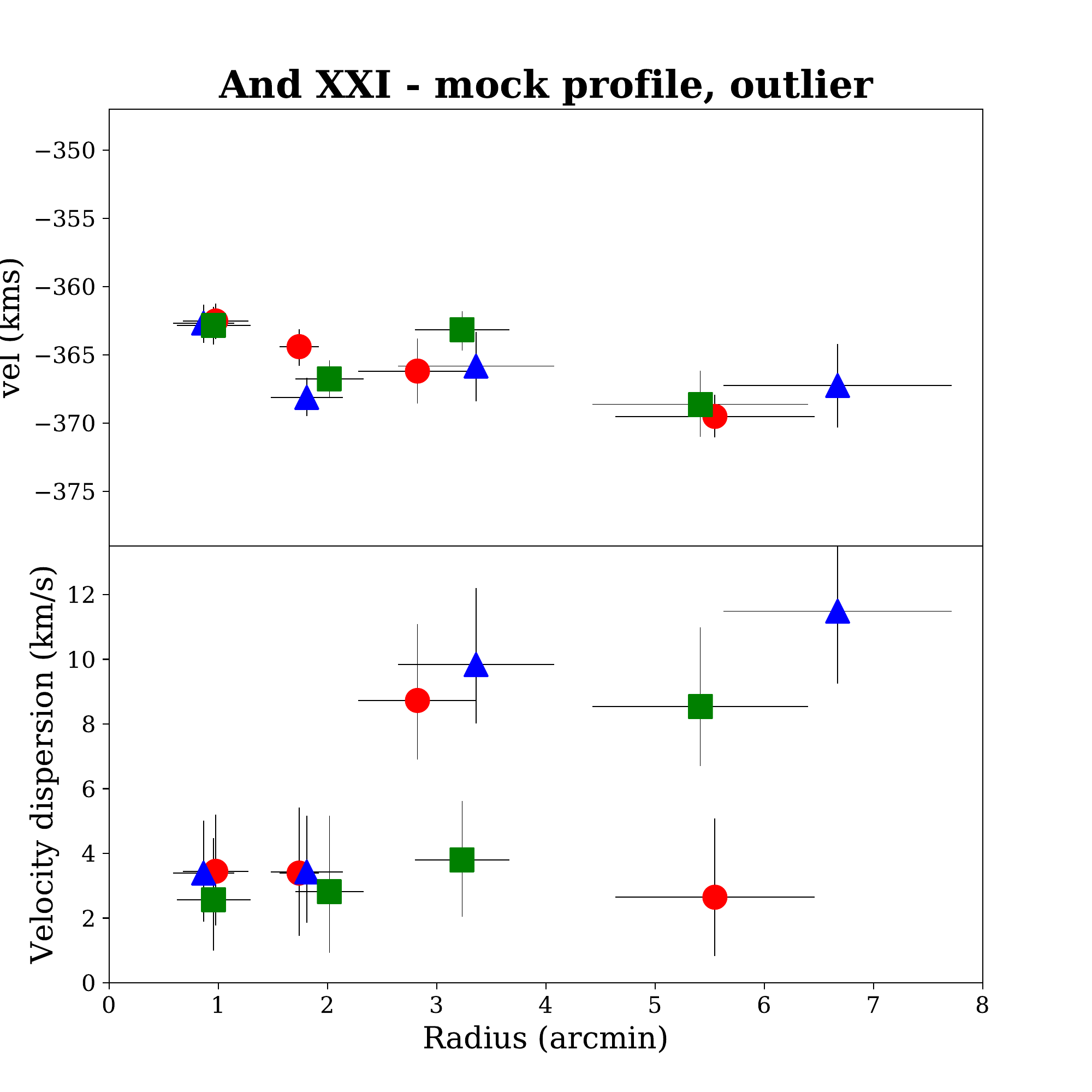}
     \caption{{\bf Left:} Radial velocity profile (top) and velocity
       dispersion profile (bottom) for 3 random realisations of our mock data (shown with red circles, green squares and blue triangles). These profiles all have chi-square values lower than our observed data, and so show `normal' flat velocity and dispersion profiles. {\bf Right:} As left, but for mock realisations where the chi-square values for both systemic velocity and velocity dispersion are similar to our observed data. We see very similar behaviour, where most of the galaxy has a dispersion of $3-5\kms$, but 1-2 bins show dispersions of order $10\kms$. This is seen in $\sim1.5\%$ of all mock realisations, indicating our observed distribution could merely be a statistical fluke. }
  \label{fig:mock}
  \end{center}
\end{figure*}

 Such bumps and wiggles in the dynamical profiles of dSphs have previously been observed (e.g. Andromeda II and XIX, \citealt{amorisco14a,collins20}). These have been interpreted as evidence of mergers, tidal effects, substructure, or just random fluctuations in the data. To investigate how often one expects to see such variation about the mean, we run tests on mock data.

First, we measure the chi-square goodness of fit our velocity and dispersion data (where we compare each bin to the mean of the full sample). We measure this to be $\chi_v=5.0$ for the velocity profile, and $\chi_s=6.4$ for the dispersion profile. Next, we generate a mock dataset for our And XXI data. We use the posteriors generated from our MCMC analysis for each of our And XXI, M31 halo and Milky Way contaminants to define the kinematics for each populations. We then randomly draw 77, 12 and 163 stars from these distributions for each population respectively (matching our observed sample sizes). These would represent the `true' velocities. We then perturb these using uncertainties equivalent to those of our observed stars. We assign them a radial position based on the observed radial distribution of our sample (which follows the shape of a log-normal profile). We then construct radial velocity profiles for this mock data in the exact same way as our real data. This process is repeated 1000 times.

We then compare the chi-square statistics for the radial profiles in $v_r$ and $\sigma_v$ for each realisation with our observations. We find that 38/1000 realisations have $\chi_v\ge5.0$, 35/1000 realisations have $\chi_s>6.4$, and 13/1000 cases satisfy both criteria. We show examples of a random selection of `normal' (better chi-square) and `outlier' (higher chi-square) realisations in fig.~\ref{fig:mock}. Statistically, our observed distribution can be drawn from a dwarf galaxy with a flat, Gaussian profile $\sim1.3\%$ of the time (a 2.5$\sigma$ event). As such, it is an interesting anomaly, but not strong evidence for substructure. More data would resolve whether this is a true feature, or merely noise in the data.

\begin{figure*}
  \begin{center}
     \includegraphics[angle=0,width=0.45\hsize]{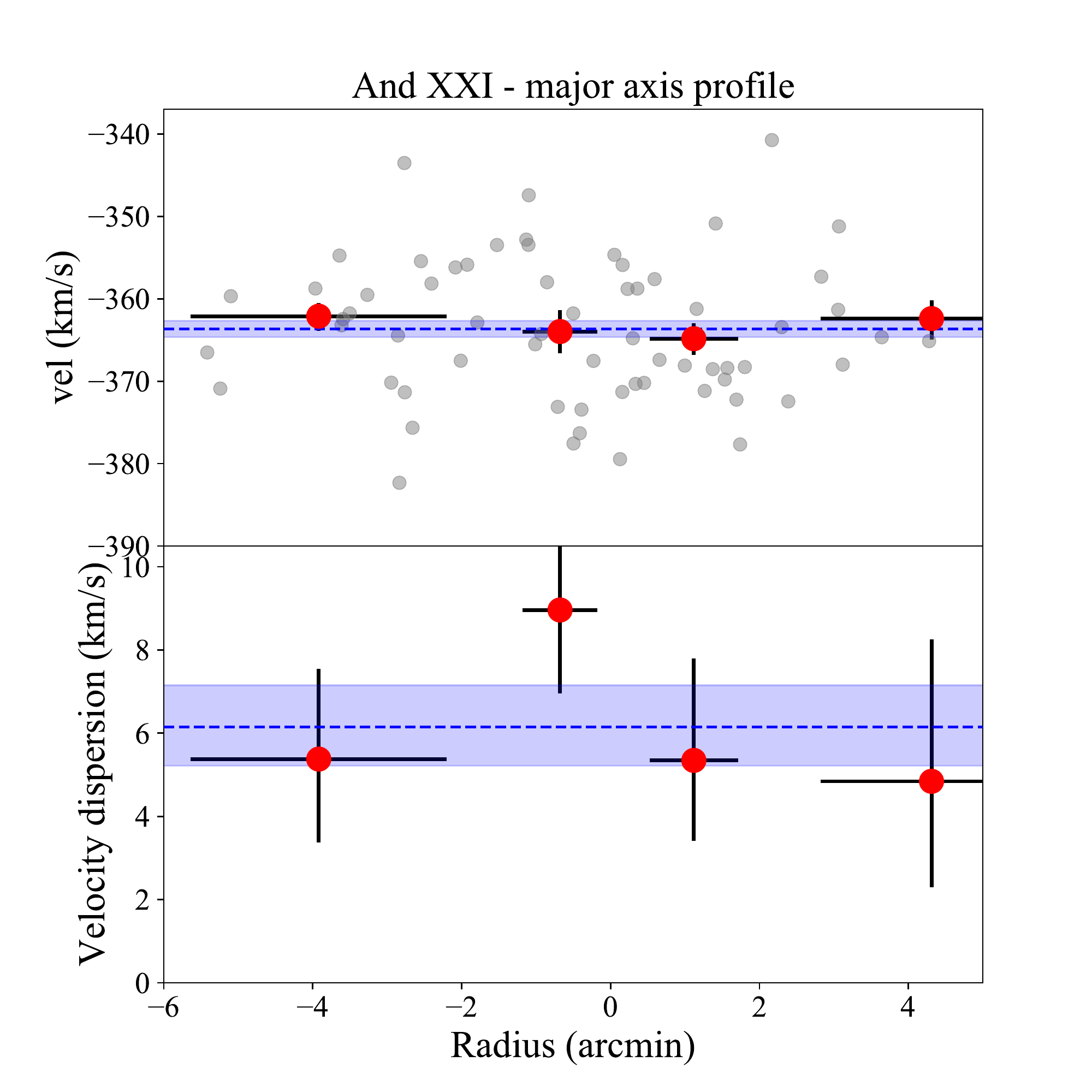}
     \includegraphics[angle=0,width=0.45\hsize]{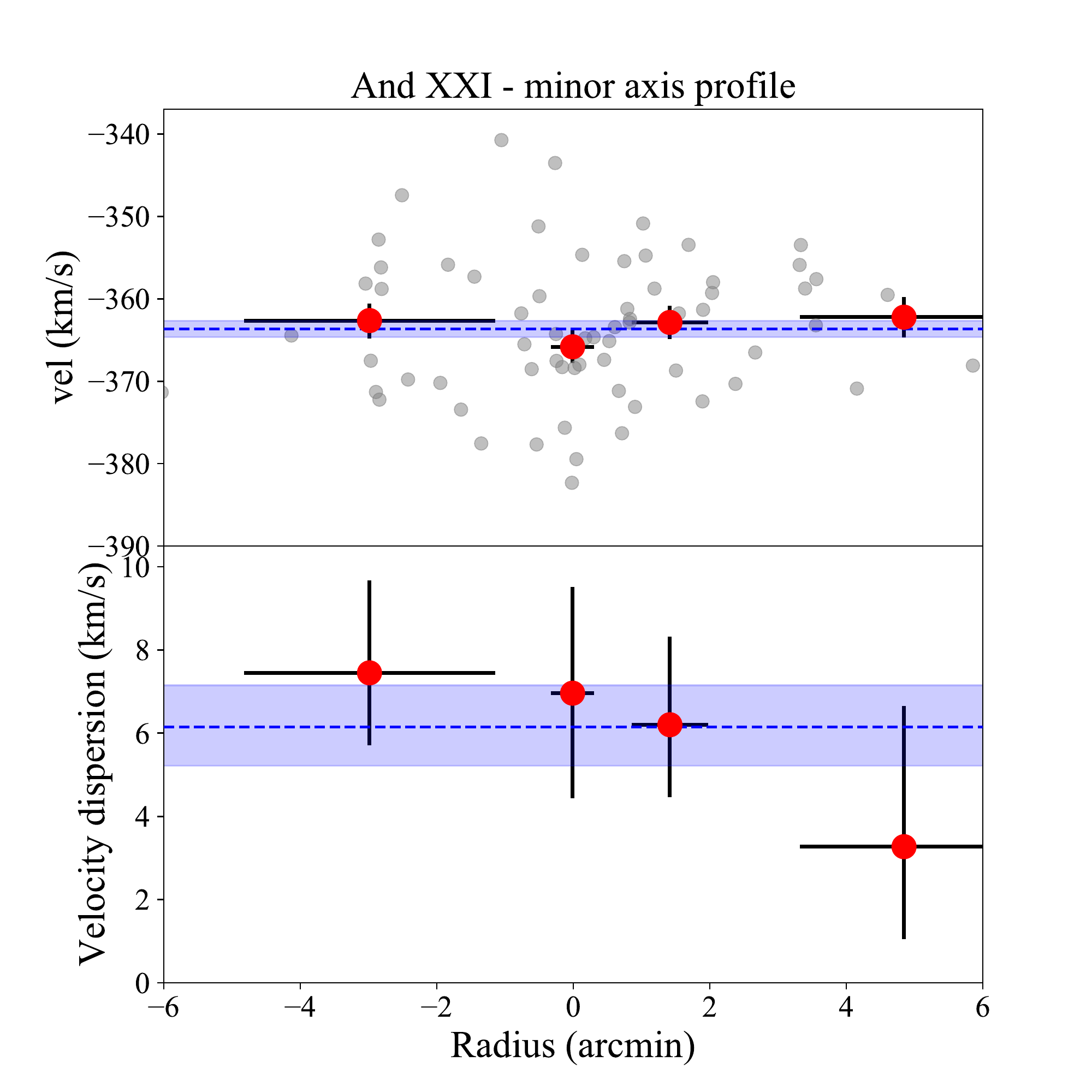}
\caption{Velocity profiles and dispersion profiles along the
   photometric major (left) and minor
  (right) axis. Neither the major
nor minor axis shows evidence for rotation or a velocity gradient, and there
is significant scatter around both the systemic velocity and average
dispersion. }
  \label{fig:axp}
  \end{center}
\end{figure*}

Finally, we construct major and minor axis profiles of And~XXI (shown in fig.~\ref{fig:axp}) to see whether there are any signs of rotation about these axes. Both appear flat, suggesting there is no (significant) rotation in And~XXI. As a final check, we use the same MCMC technique from \citet{collins20} to search for a signature of rotation along any arbitrary axis, however no such signal is found.

\subsection{ An interesting feature in the M31 halo}

Our kinematic sample also contains 13 likely M31 halo stars. There are too few to precisely determine the kinematics of the halo at this distance, but our initial MCMC process finds $v_{\rm M31}=-273^{+40}_{-37}\,\kms$ and $\sigma_{\rm M31}=135^{+28}_{-24}\,\kms$, broadly consistent with previous studies of the halo  (e.g. \citealt{chapman06,gilbert18}). But interestingly, 5 of these stars are quite clustered in velocity around $v_r=-270\,\kms$ (highlighted as grey squares in fig.~\ref{fig:vels}). 4 of these stars are also very close in CMD space, lying along an isochrone of [Fe/H]$=-1.5$ (as shown in fig.~\ref{fig:cmd}. It's possible that this is a detection of an unresolved stellar stream in the M31 halo. Features like these are often seen in M31 halo fields in both imaging and dynamics, as discussed in e.g \citet{gilbert12,gilbert18,ibata14c}, and likely relate to ancient accretions or stripping of low mass systems. 

With so few stars, it is hard to conclude much about this feature, but the frequency of such detections in both PAndAS and SPLASH is yet further evidence for the rich merger history of Andromeda.

\subsection{Metallicity}
\label{sec:feh}

\begin{figure}
  \begin{center}
       \includegraphics[angle=0,width=\columnwidth]{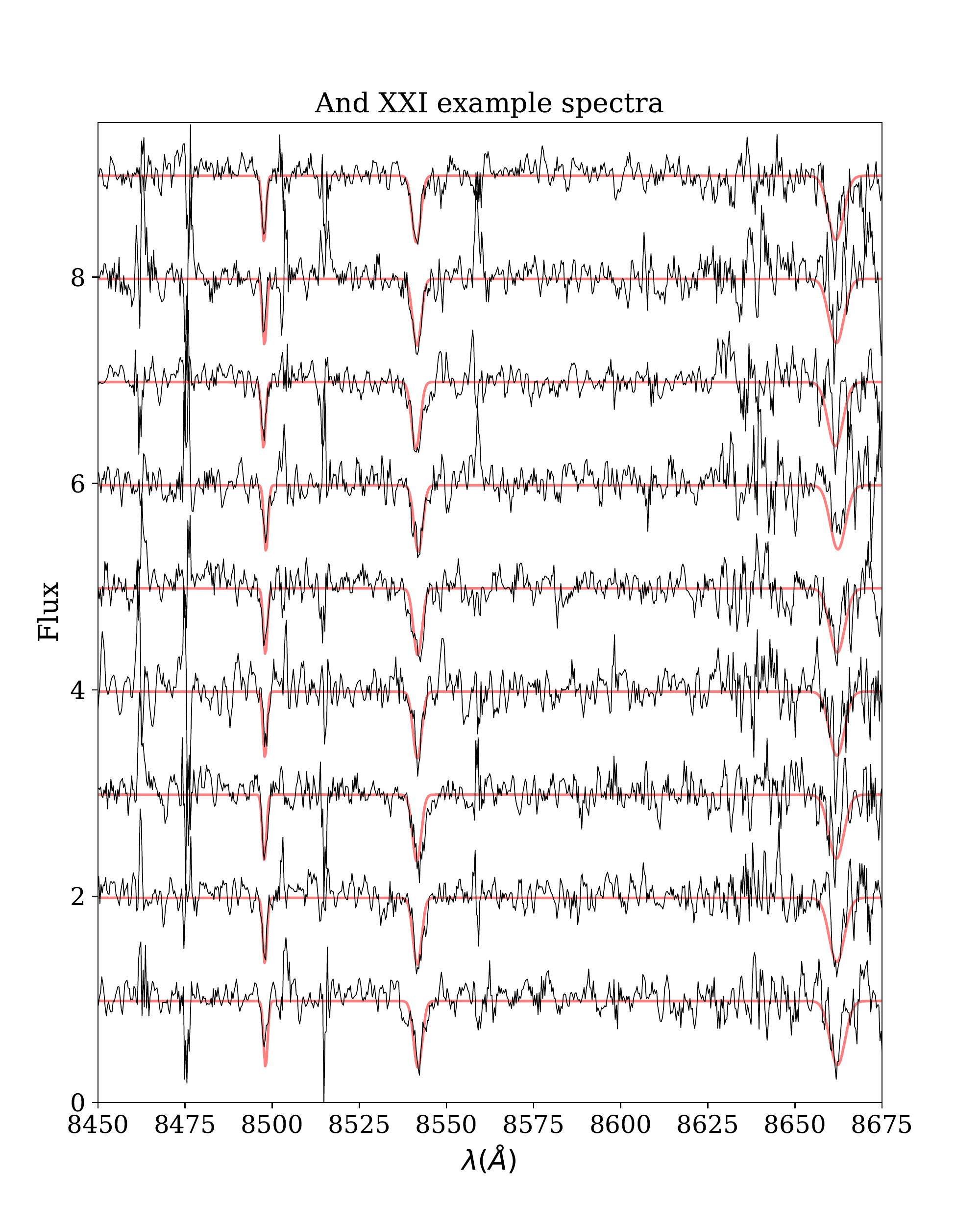}
       \caption{Continuum normalised spectra for a sample of And XXI member stars ($P_{\rm member}>0.1)$ with $S/N>5$ per pixel. The red line represents the fit to the three Ca II lines. } 
       \label{fig:spec}
  \end{center}
\end{figure}

\begin{figure*}
  \begin{center}
       \includegraphics[angle=0,width=0.47\hsize]{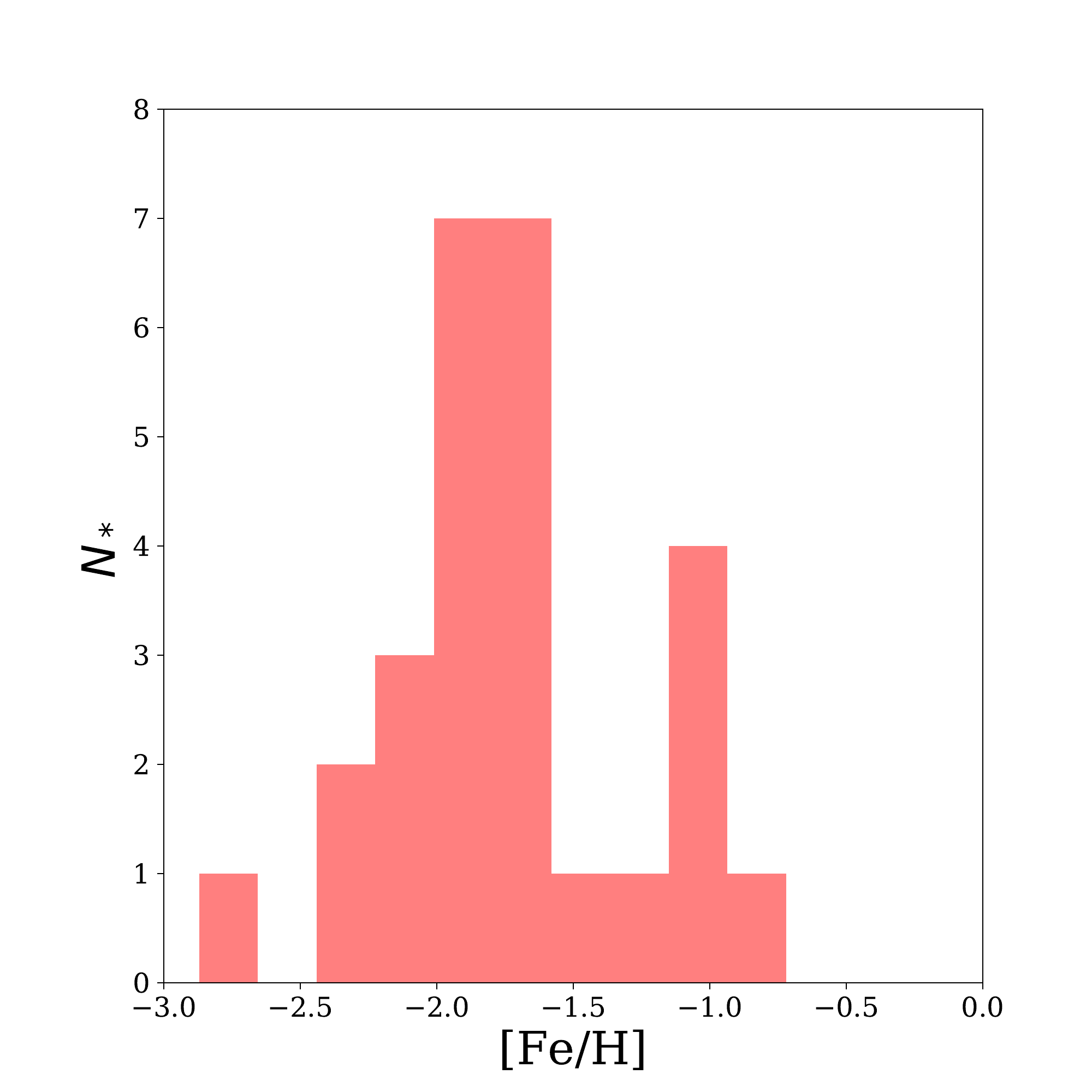}
       \includegraphics[angle=0,width=0.43\hsize]{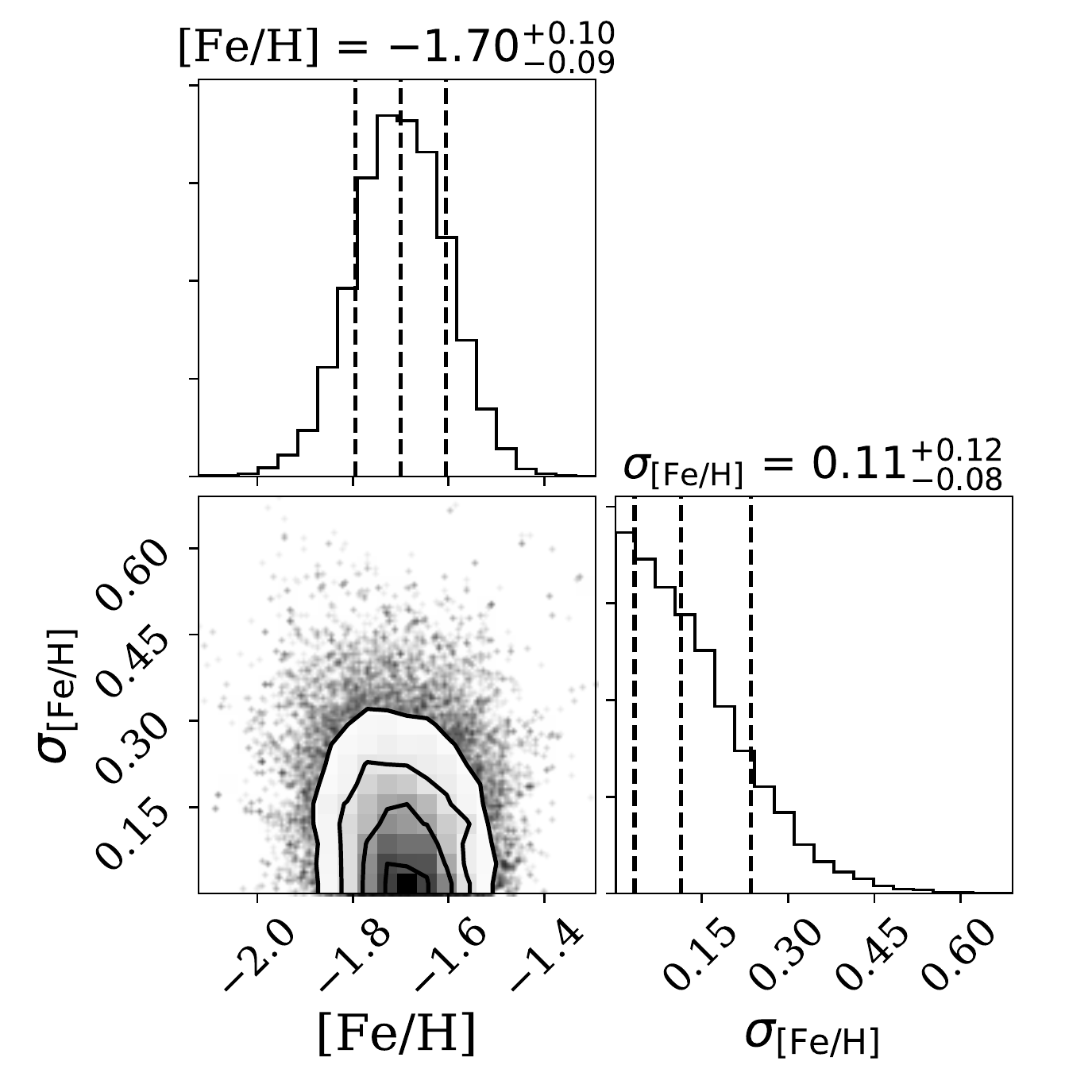}
       \caption{The metallicity distribution function for And~XXI member stars with $S/N>5$ per pixel (left), and a corner plot showing the mean and spread of the metallicity for this system (right). We find a mean [Fe/H]$=-1.7\pm0.1$ dex, and are unable to constrain the metallicity spread in the system. It is likely less than $\sigma_{\rm [Fe/H]}<0.5$ dex at 99\% confidence.} 
       \label{fig:mdf}
  \end{center}
\end{figure*}

\begin{figure}
  \begin{center}
    \includegraphics[angle=0,width=0.9\hsize]{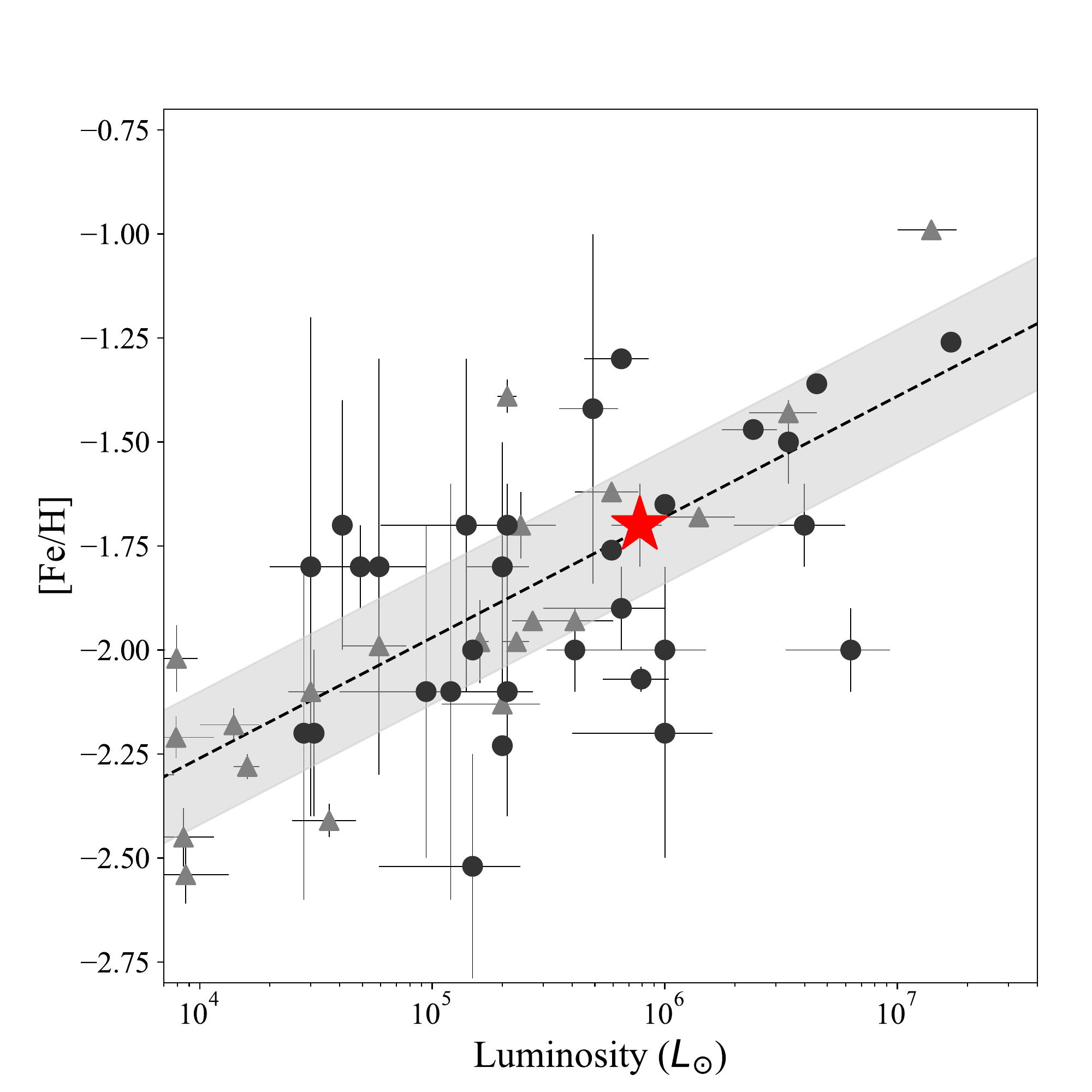}
       \caption{The luminosity-metallicity relation for Local Group dSphs. And~XXI is shown as a red star. The dSphs of M31 are represented with dark grey circles, while those of the Milky Way are light grey triangles. The black dashed line is the best fit relation for Milky Way dSphs from \citet{kirby13a}, with the grey band showing the $1\sigma$ scatter. And~XXI is perfectly consistent with this relation.} 
       \label{fig:lfeh}
  \end{center}
\end{figure}

We measure the metallicity of our And~XXI member stars using the equivalent widths of the three Ca~II triplet absorption lines. These features are well-known to allow a proxy for iron abundance measurements, [Fe/H], in RGB stars (e.g. \citealt{armandroff91}). We perform this for all stars with $S/N>5$ per pixel, and we show a representative sample of these spectra in fig.~\ref{fig:spec}. We follow the technique of \citet{collins13}, and begin by fitting the continuum of each spectrum, and then normalising them (such that the mean continuum is equal to one). We then fit the three Ca II lines with Gaussian profiles, and use the calibration of \citet{starkenburg10} to convert their equivalent widths to a measure of [Fe/H]. We present the metallicity distribution function for 30 stars that pass the $S/N$ cut, and that have metallicity uncertainties of less than 0.8 dex in fig.~\ref{fig:mdf}. We see that the distribution is centred around $[{\rm Fe/H}]=-1.7\pm0.1$. We cannot resolve a metallicity spread given the large uncertainties in our measurements,  measuring $\sigma_{\rm [Fe/H]}=0.1\pm0.1$ dex. We find $\sigma_{\rm [Fe/H]}<0.5$ dex at a 99\% confidence. There is perhaps an interesting over density of stars with [Fe/H]$\sim-1$, which could imply a relatively metal rich sub-population and a more complex SFH than the single orbit HST imaging of And~XXI imply \citep{weisz19a}. The mean metallicity for And~XXI is perfectly consistent with the luminosity-metallicity relation for Local Group dwarf galaxies, as can be seen in fig.~\ref{fig:lfeh} (with the Andromeda subsystem shown as red circles, and the Milky Way as blue triangles, \citealt{kirby13a,tollerud12,collins13,collins17,collins20,kirby13a,martin13a,ho15,wojno20}). The best fit to this relation from \citet{kirby13a} is shown as a dashed black line, with the grey band representing the $1\sigma$ scatter. 

\section{The mass profile of And~XXI}
\label{sec:mass}

Using our derived velocity dispersion, we can measure the central mass, density and mass-to-light ratio of And~XXI, and compare it with similar dwarf galaxies in the Local Group. It has been well-established that one can constrain the mass within a given radius, $R$, for dwarf spheroidals using their velocity dispersions. Typically, $R$ is similar or equal to the projected half light radius of the galaxy \citep{walker09b,wolf10}. Recent work by \citet{errani18} use the mass within $R=1.8r_{\rm half}$, and no assumption on the shape of the mass (dispersion) profile, such that:

\begin{equation}
M(R<1.8\,r_{\rm half})=\frac{3.5r_{\rm half}\sigma_v^2}{G}.
\end{equation}

\begin{figure*}
  \begin{center}
     \includegraphics[angle=0,width=0.45\hsize]{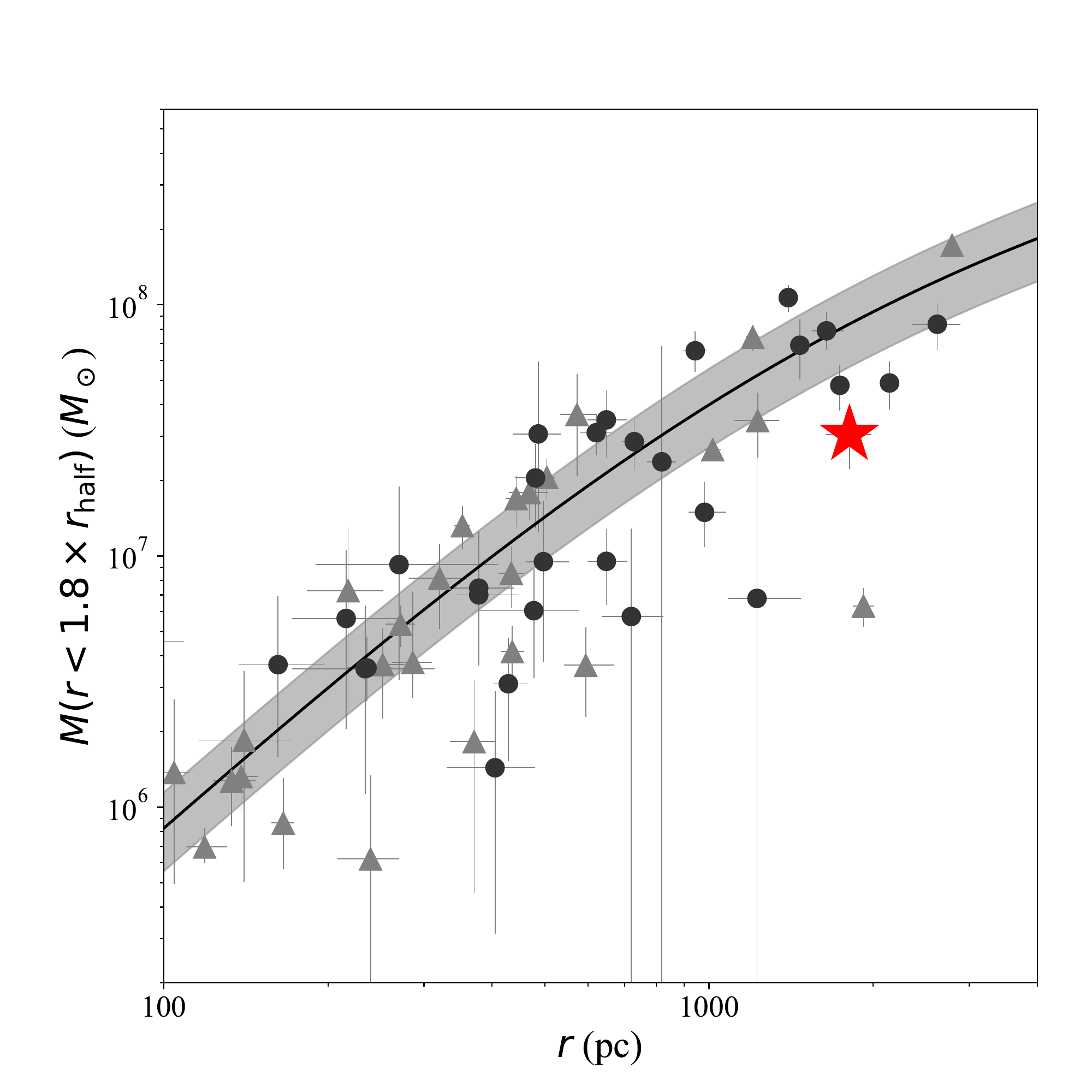}
     \includegraphics[angle=0,width=0.45\hsize]{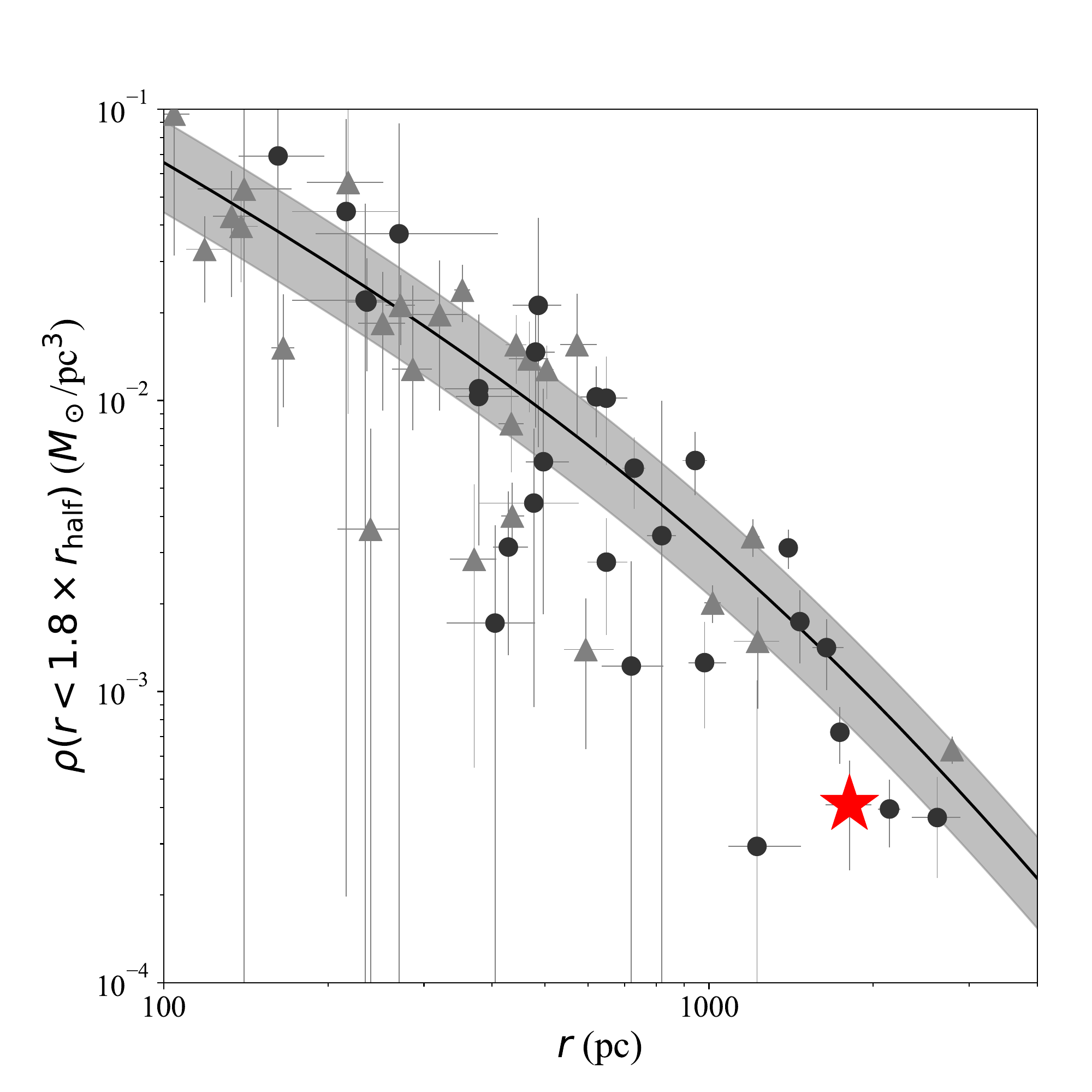}
      \caption{{\bf Left:} Mass enclosed within $1.8\times r_{\rm half}$ for the Local Group dSphs. The MW subsystem are shown with light grey triangles, while the M31 system are shown as dark grey circles. And~XXI is highlighted as a red star. The black line shows the best fit  NFW halo profile for the Local Group dSphs from \citet{collins14}, with the gray band representing the $1\sigma$ scatter, And~XXI is significantly below this fit, indicating it has a lower mass than expected for its size. {\bf Right:} Enclosed density for the Local Group dSphs, plus the best fit density profile, as in the left-hand panel. And~XXI is lower density than typically expected.}
  \label{fig:summary}
  \end{center}
\end{figure*}

Using this, we measure a central mass of $M(R<1.8\,r_ {\rm half})=3.0^{+0.9}_{-0.8} \times10^7\,{\rm M}_\odot$. This gives a central mass-to-light ratio  of $[M/L](R<1.8\,r_ {\rm half})=78^{+30}_{-28}\,{\rm M}_\odot/{\rm L}_\odot$, significantly dark matter dominated. We measure a central dark matter density of $\rho(R<1.8\,r_ {\rm half})=4.1\pm1.7\times10^{-4}\,{\rm M}_\odot\,{\rm pc}^{-3}$. In our previous work, we found that And~XXI had a lower central mass and density than dwarf galaxies of a comparable size or brightness. With our updated measurements, we find this remains true. In fig.~\ref{fig:summary}, we show the mass and density within $1.8\,r_{\rm half}$ for Milky Way (blue triangles) and M31 (red circles) dSphs, compared to the best fit NFW mass(/density) profiles for this population from \citet{collins14} in grey  \citep{walker07,walker09b,simon07,simon11,simon15,martin07,martin13a,martin14b,ho12,tollerud12,collins13,collins20,kirby15a,kirby17b}. In both panels, we see that And~XXI is of significantly lower mass and density than the best fit profile would predict. Could it be lower density because it harbours a central core, as is supposed for some low-density Milky Way dSphs (e.g. Fornax, Crater 2 and Antlia 2 \citep{goerdt06,walker11,amorisco12,read19a,caldwell17,torrealba19a}? And is this growing number of low density dwarf galaxies a challenge for the cold dark matter paradigm? To test this, we use Jeans modelling to measure the density profile of And~XXI, and compare to $\Lambda{\rm CDM}$ expectations.

\subsection{Mass modelling with \GravSphere}\label{sec:gravsphere}

We use an updated version of the {\sc GravSphere} Jeans modelling code\footnote{A version of the {\sc GravSphere} code with the free form mass model, amongst others \citep{genina20}, is available to download from \href{https://github.com/AnnaGenina/pyGravSphere}{https://github.com/AnnaGenina/pyGravSphere}. The updated {\sc GravSphere} code described in this paper, along with the new {\sc binulator} binning method (\S\ref{sec:binulator}), is available to download from \href{https://github.com/justinread/gravsphere}{https://github.com/justinread/gravsphere}.} to model our dynamical and photometric data for And~XXI, with the end goal of measuring its dark matter density profile, $\rho(r)$. {\sc GravSphere} is described in detail in \citet{read17b} and \citet{read18a}; here we briefly summarise its implementation. {\sc GravSphere} solves the spherical Jean equation \citet{jeans22} for a set of `tracers' (i.e. our member stars with radial velocity measurements), to determine both $\rho(r)$, and the velocity anisotropy profile, $\beta(r)$. The Jeans equation is given by \citep{vandermarel94, mamon05}:

\begin{equation}
\sigma_{\rm los}^2(R)=\frac{2}{\Sigma(R)}\int_R^\infty \left(1-\beta(r)\frac{R^2}{r^2}\right) \nu(r)\sigma_r^2\frac{r{\rm d}r}{\sqrt{r^2-R^2}},
\label{eq:jeans}
\end{equation}

\noindent where $\Sigma(R)$ is the surface brightness profile at projected radius $R$, $\nu(r)$ is the spherically averaged tracer density as a function of spherical radius, $r$, and $\beta(r)$ is the velocity anisotropy:

\begin{equation}
\beta=1-\frac{\sigma_t^2}{\sigma_r^2}.
\label{eq:aniso}
\end{equation}
where $\sigma_r$ and $\sigma_t$ are the radial and tangential velocity dispersion profiles, respectively, and $\sigma_r$ is given by:

\begin{equation}
\sigma_r^2(r)=\frac{1}{\nu(r)g(r)}\int_r^\infty \frac{G M({\tilde r})\nu({\tilde r})}{{\tilde r}^2}g({\tilde r}){\rm d}r,
\label{eq:sigr}
\end{equation}
where:

\begin{equation}
g(r)= \exp\left(2\int\frac{\beta(r)}{r} \dd r \right),
\label{eqn:gr}
\end{equation}
and $M(r)$ is the cumulative mass profile of the system.

The light profile, $\Sigma(R)$, is modelled as a superposition of Plummer spheres \citep{plummer11,rojasnino16}. The cumulative mass profile is given by:

\begin{equation}
    M(r) = M_*(r) + M_{\rm cNFWt}(r) ,
\end{equation}
where the cumulative stellar light profile, $M_*(r)$, is normalised to asymptote to the total stellar mass at infinity, and allowed to vary within some flat prior range (see \S\ref{sec:priors}), and $M_{\rm cNFWt}(<r)$ is the \coreNFWtides\ profile from \citet{read18a} that describes the cumulative dark matter mass profile:

\begin{equation}
M_{\rm cNFWt}(r) =
\left\{
\begin{array}{ll}
  M_{\rm cNFW}(<r) & r < r_t  \vspace{4mm}\\
M_{\rm cNFW}(r_t) \,\, + & \\
4\pi \rho_{\rm cNFW}(r_t) \frac{r_t^3}{3-\delta}
\left[\left(\frac{r}{r_t}\right)^{3-\delta}-1\right] & r > r_t
\end{array}
\right.
\label{eqn:McNFWt}
\end{equation}
where:
\begin{equation}
M_{\rm cNFW}(<r) = M_{\rm NFW}(<r) f^n
\label{eqn:coreNFW}
\end{equation}
and:

\begin{equation} 
M_{\rm NFW}(r) = M_{200} g_c \left[\ln\left(1+\frac{r}{r_s}\right) - \frac{r}{r_s}\left(1 + \frac{r}{r_s}\right)^{-1}\right]
\label{eqn:MNFW}
\end{equation}
with:

\begin{equation} 
f^n = \left[\tanh\left(\frac{r}{r_c}\right)\right]^n
\end{equation}
and $\rho_{\rm cNFW}$ is given by:
\begin{equation} 
\rho_{\rm cNFW}(r) = f^n \rho_{\rm NFW} + \frac{n f^{n-1} (1-f^2)}{4\pi r^2 r_c} M_{\rm NFW}
\label{eqn:rhocNFW}
\end{equation}
where:
\begin{equation} 
\rho_{\rm NFW}(r) = \rho_0 \left(\frac{r}{r_s}\right)^{-1}\left(1 + \frac{r}{r_s}\right)^{-2}
\label{eqn:rhoNFW}
\end{equation}
and the central density $\rho_0$ and scale length $r_s$ are given by: 
\begin{equation} 
\rho_0 = \rho_{\rm crit} \Delta c_{200}^3 g_c / 3 \,\,\,\, ; \,\,\,\, r_s = r_{200} / c_{200}
\end{equation}
\begin{equation}
g_c = \frac{1}{{\rm log}\left(1+c_{200}\right)-\frac{c_{200}}{1+c_{200}}}
\end{equation}
and
\begin{equation} 
r_{200} = \left[\frac{3}{4} M_{200} \frac{1}{\pi \Delta \rho_{\rm crit}}\right]^{1/3}
\label{eqn:r200}
\end{equation} 
where $c_{200}$ is the dimensionless {\it concentration parameter}; $\Delta = 200$ is the over-density parameter; $\rho_{\rm crit} = 136.05$\,M$_\odot$\,kpc$^{-3}$ is the critical density of the Universe at redshift $z=0$; $r_{200}$ is the `virial' radius at which the mean enclosed density is $\Delta \times \rho_{\rm crit}$; and $M_{200}$ is the `virial' mass -- the mass within $r_{200}$. The \coreNFWtides\ profile adds to these a parameter $n$ that determines how centrally cusped the density is ($n=0$ corresponds to a central constant density core; $n=1$ to a $\rho \propto r^{-1}$ cusp), $r_c$ that sets the size of this core, and $r_t$ that determines an outer `tidal radius' beyond which the density fall-off steepens as $\rho \propto r^{-\delta}$.

The velocity anisotropy profile is given by: 
\begin{equation} 
\beta(r) = \beta_0 + \left(\beta_\infty-\beta_0\right)\frac{1}{1 + \left(\frac{r_0}{r}\right)^q}
\label{eqn:beta}
\end{equation}
which makes the solution to equation \ref{eqn:gr} analytic. 

{\sc GravSphere} fits the surface brightness profile, $\Sigma(R)$, and radial velocity dispersion profile, $\sigma_{\rm los}$, using the Markov chain Monte Carlo code {\sc Emcee} \citep{fm13a}. A symmeterised version of $\beta(r)$ is used to avoid infinities, defined as:

\begin{equation}
{\tilde \beta} = \frac{\sigma_r^2-\sigma_t^2}{\sigma_r^2+\sigma_t^2}=\frac{\beta}{2-\beta},
\end{equation}

\noindent where ${\tilde \beta}=0$ describes an isotropic velocity distribution, ${\tilde \beta}=-1$ a fully tangential distribution, and ${\tilde \beta}=1$ a fully radial distribution. {\sc GravSphere} also fits two `virial shape parameters' to break the well-known degeneracy between $\rho$ and $\beta$ \citep{merrifield90,richardson14,read17b}.

{\sc GravSphere} has been extensively tested on mock data for spherical and triaxial systems \citep{read17a,read21}, for tidally disrupting mocks \citep{read18a} and for realistic mocks drawn from a cosmological simulation \citep{genina20}. In most cases, \GravSphere\ is able to recover the dark matter density profile within its 95\% confidence intervals over the range $0.5 < R/R_{1/2} < 2$, where $R_{1/2}$ is the half light radius. However, the code has a tendency to underestimate the density at large radii $R \gtrsim 2R_{1/2}$ \citep{read17a,read21}. Furthermore, for small numbers of stars, or where the measurement error on each star is large, the binning method within the code can become biased \citep{gregory19a,Zoutendijk21}.

To address the above concerns, for this paper we have extensively updated and improved {\sc GravSphere}. The primary change is a new data binning module, {\sc binulator}, that we describe in \S\ref{sec:binulator}. Smaller additional changes are: 1) a switch to using the \coreNFWtides\ profile as the default dark matter mass model, as in \citet{read18a,read19b}, rather than using the `non-parametric' series of power laws centred on a set of radial bins, described in \citet{read17b}; and 2) using tighter priors on $\beta(r)$ by default. We describe all of our priors in \S\ref{sec:priors} and present mock data tests of our new methodology in \S\ref{app:mocks}. 

The reasons for shifting away from the `non-parametric' dark matter mass profile from \citet{read17b} is twofold. Firstly, for ${\sim}1000+$ tracers, the results from using the \coreNFWtides\ profile and the `non-parametric' profile agree within their respective 68\% confidence intervals \citet{alvarez20}, yet the \coreNFWtides\ model fit parameters are more cosmologically useful \citep{read18a,read19b}. Secondly, in the absence of data, the default priors on the non-parametric mass profile favour a high density in the centre that falls very steeply outwards, inconsistent with expectations in most popular cosmological models. By contrast, the \coreNFWtides\ model naturally defaults to cosmological expectations at large radii in the absence of data.

The reason for the tighter priors on $\tilde{\beta}(r)$ is that, theoretically, we expect that dynamical systems in pseudo-equilibrium should be close to isotropic in the centre, with radial anisotropy, or weak tangential anisotropy, at large radii \citep[e.g.][]{pontzen15,genina20,alvey21}. For this reason, we allow for only mild tangential anisotropy ($\tilde{\beta}(r) > -0.1$) and demand $\tilde{\beta}(r) \rightarrow 0$ for $r\rightarrow 0$. This prior is also consistent with measurements of $\tilde{\beta}(r)$ in Milky Way satellites such as Draco and Sculptor (e.g. \citet{massari18,massari20}).

\subsubsection{The {\sc binulator}}\label{sec:binulator}

The main improvement we make to {\sc GravSphere} is a complete reworking of its data binning routines into a separate code: the {\sc binulator}. This improves the binning by fitting a generalised Gaussian probability distribution function (PDF) to each bin to estimate its mean, variance and kurtosis. This has the advantages that: 1) the distribution function can be readily convolved with the error PDF of each star, survey selection functions, binary star velocity PDFs, and similar; and 2) the method returns a robust estimate of the mean, variance and kurtosis, and their uncertainties, even in the limit of a very small number of tracer stars.

The full method proceeds, as follows. Firstly, the data for $\Sigma(R)$ are fit using {\sc Emcee} \citep{fm13a} to obtain the best-fit multi-Plummer model for the light profile (see above). This provides an initial guess for the later \GravSphere\ fits, and will be important for calculating the virial shape parameters (see below). To provide maximum flexibility, these fits allow for individual Plummer components to have `negative mass' while still ensuring that the total surface density is positive definite, as in \citet{rojasnino16}. Next, the discreet stellar velocity data are sorted into equal number bins in radius, weighted by membership probability. We use 25 stars per bin by default; our results are not sensitive to this choice. A velocity PDF is then fit to the stars in each bin to determine the mean, variance and kurtosis of the bin, similarly to the method described in \citet{sanders20}. However, for our velocity PDF, we assume a generalised Gaussian:

\begin{equation}
p_i = \frac{\beta_v}{2\alpha_v\Gamma(1.0/\beta_v)}\exp\left(-(|v_{{\rm los},i}-\mu_v|/\alpha_v)^\beta_v)\right)
\label{eqn:normgaus}
\end{equation}
where $v_{{\rm los},i}$ is the line of sight velocity of a star, $i$, $\Gamma(x)$ is the Gamma Function, and $\mu_v$, $\alpha_v$ and $\beta_v$ are parameters fit to each bin that relate to moments of the velocity distribution function. The mean is given by $\mu_v$; the variance by $\sigma_{\rm los}^2 = \alpha_v^2\Gamma(3.0/\beta_v)/\Gamma(1.0/\beta_v)$; and the kurtosis by $\kappa = \Gamma(5.0/\beta_v)\Gamma(1.0/\beta_v)/[\Gamma(3.0/\beta_v)]^2$.

\begin{figure}
    \centering
    \includegraphics[width=0.45\textwidth]{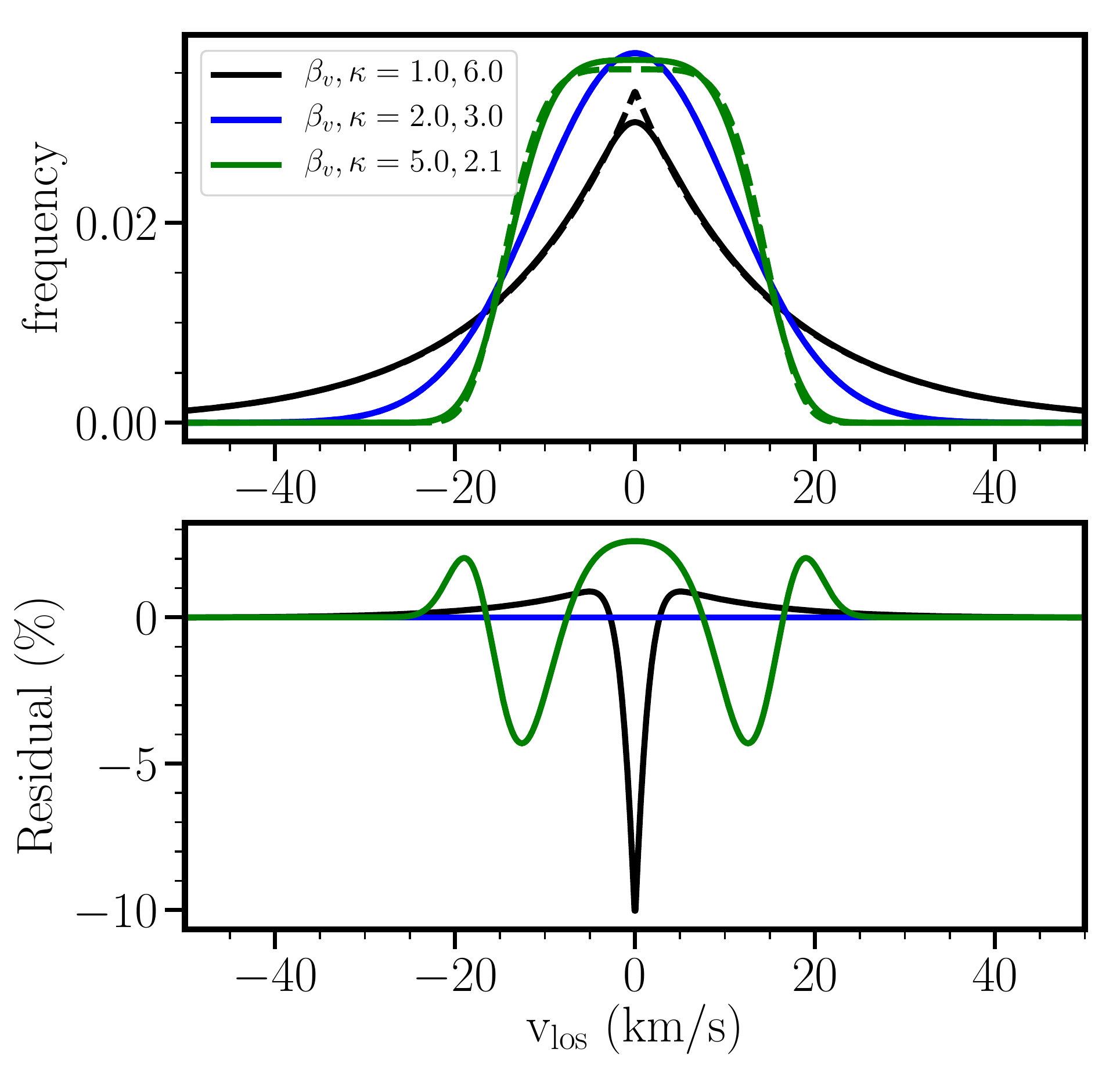}
    \caption{The generalised Gaussian velocity PDF (equation \ref{eqn:normgaus}), convolved with a Gaussian error PDF of width $\sigma_{{\rm e}} = 2$\,km/s. The top panel shows example PDFs for $\alpha_v = 15$\,km/s, with varying $\beta_v$, as marked. The corresponding kurtosis, $\kappa$, for each model is marked in the legend. Notice that $\beta_v = 2$ corresponds to a Gaussian PDF. The solid lines show the exact convolved PDF; the dashed lines show the fast analytic approximation in equation \ref{eqn:normgausapprox}. The bottom panel shows the percentage residuals between the exact and fast PDFs, defined as $({\rm PDFtrue} - {\rm PDFfast})/{\rm max[PDFtrue]} \times 100$. Notice that the error in this approximation is typically less than 5\%, and everywhere less than 10\%. The normalised Gaussian PDF allows us to measure the kurtosis in each bin, capturing distributions that are both more flat-topped and peakier than Gaussian. This is important for determining the virial shape parameters  (\S\ref{sec:binulator}), using these to break the $\rho-\beta$ degeneracy.}
    \label{fig:normgaus}
\end{figure}

To account for errors on the velocity of each star, the above generalised Gaussian should each be convolved with the error probability distribution function (PDF) for each star. This convolution integral, however, is expensive to compute. To speed up the calculation, we employ an analytic approximation to this convolution integral that assumes Gaussian errors for the stellar velocities and is exact in the limits that: 1) the normalised Gaussian approaches a Gaussian ($\beta_v = 2$); and/or 2) in the limit that the individual error on each star approaches zero; and/or 3) in the limit that the error on each star is large as compared to the variance of the velocity PDF:

\begin{equation}
    \alpha^2 \rightarrow \tilde{\alpha}^2 = \alpha^2+\sigma_{{\rm e},i}^2 \Gamma(1.0/\beta_v)/\Gamma(3.0/\beta_v)
    \label{eqn:normgausapprox}
\end{equation}
where $\sigma_{{\rm e},i}$ is the Gaussian width of the error PDF of star, $i$. The quality of this approximation is shown in Figure \ref{fig:normgaus}, along with some example generalised Gaussian PDFs. Notice that for this example (typical of current data for nearby dwarf spheroidals), the error in this approximation is typically less than 5\%, and everywhere less than 10\%.

We fit the above normalised Gaussian to each bin with {\sc Emcee} using the likelihood: 

\begin{equation}
    \mathcal{L} = \prod_i^N p_i
\end{equation}
where $N$ is the (weighted) number of stars in each bin.

We assume flat priors on the parameters: $-0.1 < \mu_v / {\rm km s}^{-1} < 0.1$; $1 < \alpha_v/ {\rm km s}^{-1} < 25$; and $1 < \beta_v < 5$. This allows for a kurtosis in the range $2 < \kappa < 6$. 

The {\sc binulator} returns marginalised PDFs of $\langle v_{\rm los}\rangle$, $\sigma_{\rm los}$ and $\kappa$ for each bin. From these, we can also calculate the marginalised PDF of fourth velocity moments $\langle v_{\rm los}^4 \rangle = \kappa \sigma_{\rm los}^4$. We use this, along with the fit to $\Sigma(R)$, to determine the marginalised PDFs of the first and second virial shape parameters:

\begin{equation} 
\vsone = \int_0^{\infty} \Sigma \vlosfour\, R\, \dd R
\label{eqn:vs1data}
\end{equation}
and
\begin{equation} 
\vstwo = \int_0^{\infty} \Sigma \vlosfour\, R^3\, \dd R \ .
\label{eqn:vs2data}
\end{equation}
where the above integrals are calculated numerically for 2,500 random draws from the marginalised distribution of $\vlosfour$ in each bin, assuming that $\vlosfour$ is constant beyond the outermost bin. We find that the marginalised PDFs for $\sigma_{\rm los}$ for each bin are close to Gaussian. For this reason, \GravSphere\ assumes Gaussian uncertainties on $\sigma_{\rm los}$ when performing its fit to $\sigma_{\rm los}(R)$. However, the PDFs for $\vsone$ and $\vstwo$ are typically non-Gaussian and so a final improvement is that \GravSphere\ now incorporates these non-Gaussian PDFs for $\vsone$ and $\vstwo$ self-consistently into its likelihood function.

Since we now fit a velocity PDF to each bin, the {\sc binulator} will work equally well with very low numbers of tracers stars and/or when the velocity errors are large, resolving the issues with earlier versions of \GravSphere, outlined above. We present tests of this updated {\sc binulator}+\GravSphere\ code on mock data in Appendix \ref{app:mocks}.

\subsubsection{Priors}\label{sec:priors}

We use priors on the \coreNFWtides\ model parameters of: $7.5 < \log_{10}(M_{200}/{\rm M}_\odot) < 11.5$; $7 < c_{200} < 53$; $-2 < \log_{10}(r_c/{\rm kpc}) < 1$; $1 < \log_{10}(r_t/{\rm kpc}) < 20$; $3 < \delta < 5$; and $-1 < n < 1$ (where $n=-1$ is steeper than a NFW cusp, and $n=1$ corresponds to a core). We use priors on the symmetrised velocity anisotropy of: $\tilde{\beta}(0) = 0$; $-0.1 < \tilde{\beta}_{\infty} < 1$; $-2 < \log_{10}(r_0/{\rm kpc}) < 0$; and $1 < q < 3$. Finally, we use a flat prior on the stellar mass of $5.7 < M_*/(10^5 M_\odot) < 9.5$ \citep{mcconnachie12}. Our results are not sensitive to these choices.

\subsubsection{{\sc GravSphere} modelling of And~XXI}

\begin{figure*}
  \begin{center}
     \includegraphics[angle=0,width=0.45\hsize]{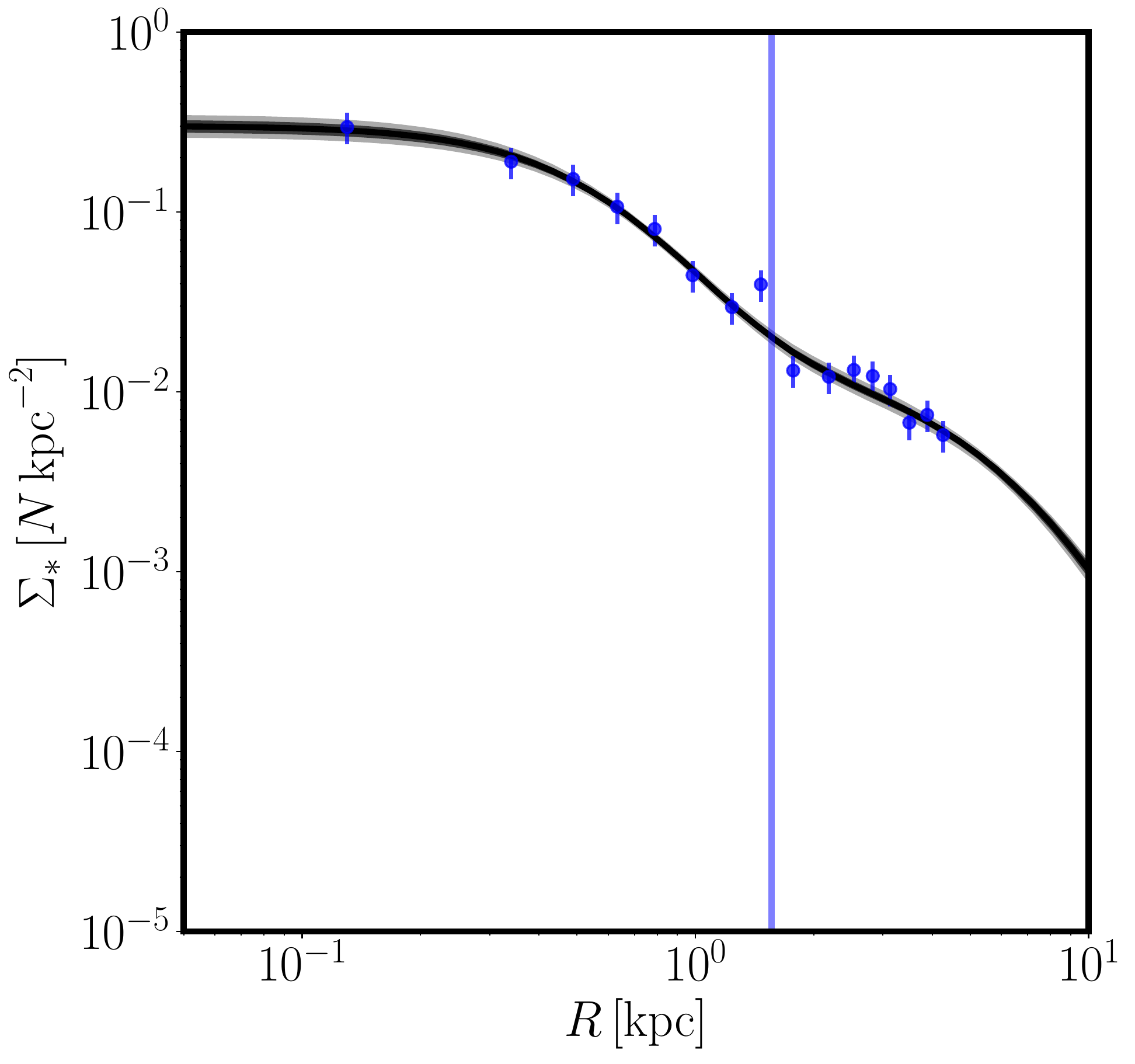}
     \includegraphics[angle=0,width=0.435\hsize]{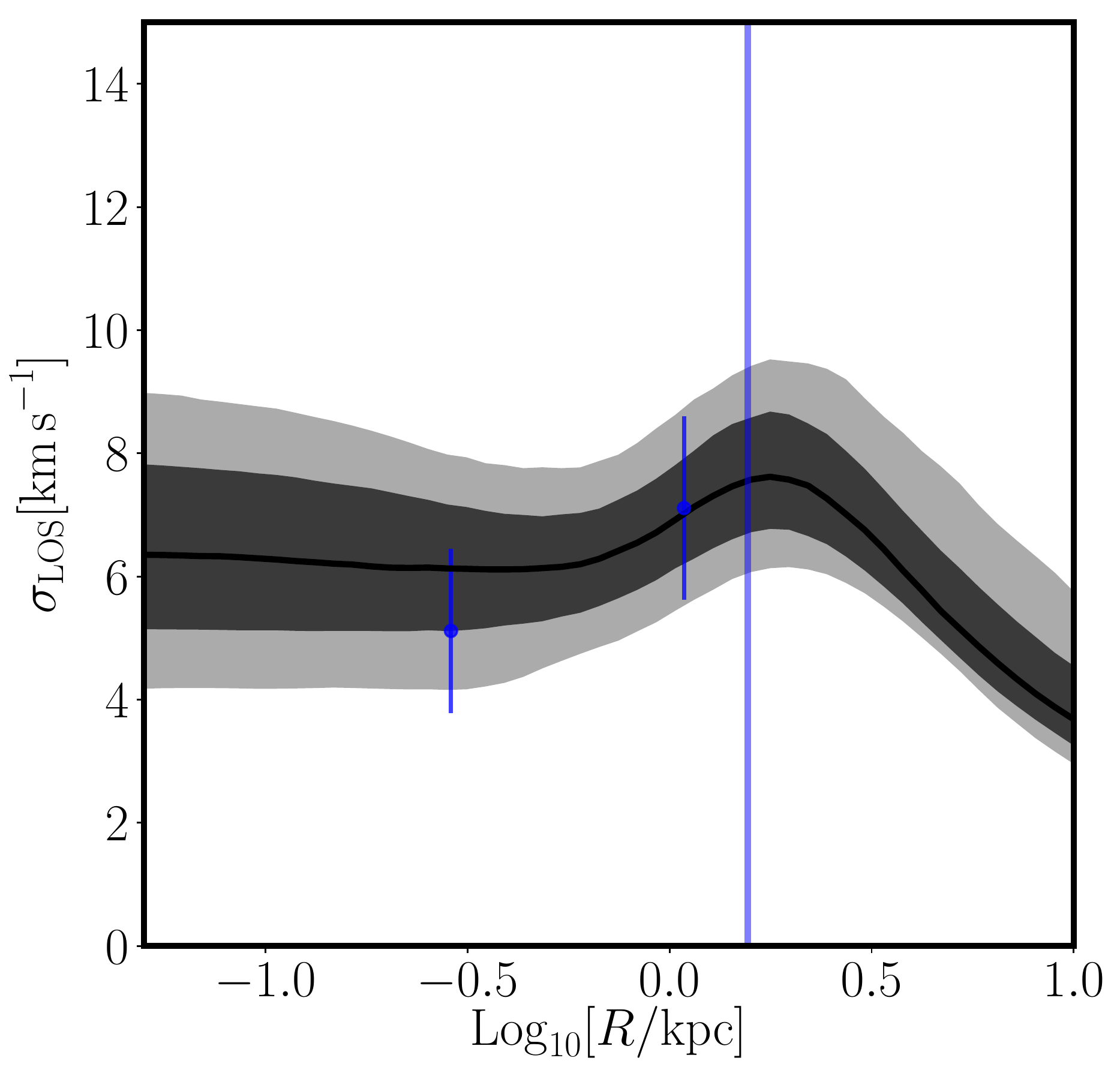}\\
    \hspace{-3mm}\includegraphics[angle=0,width=0.465\hsize]{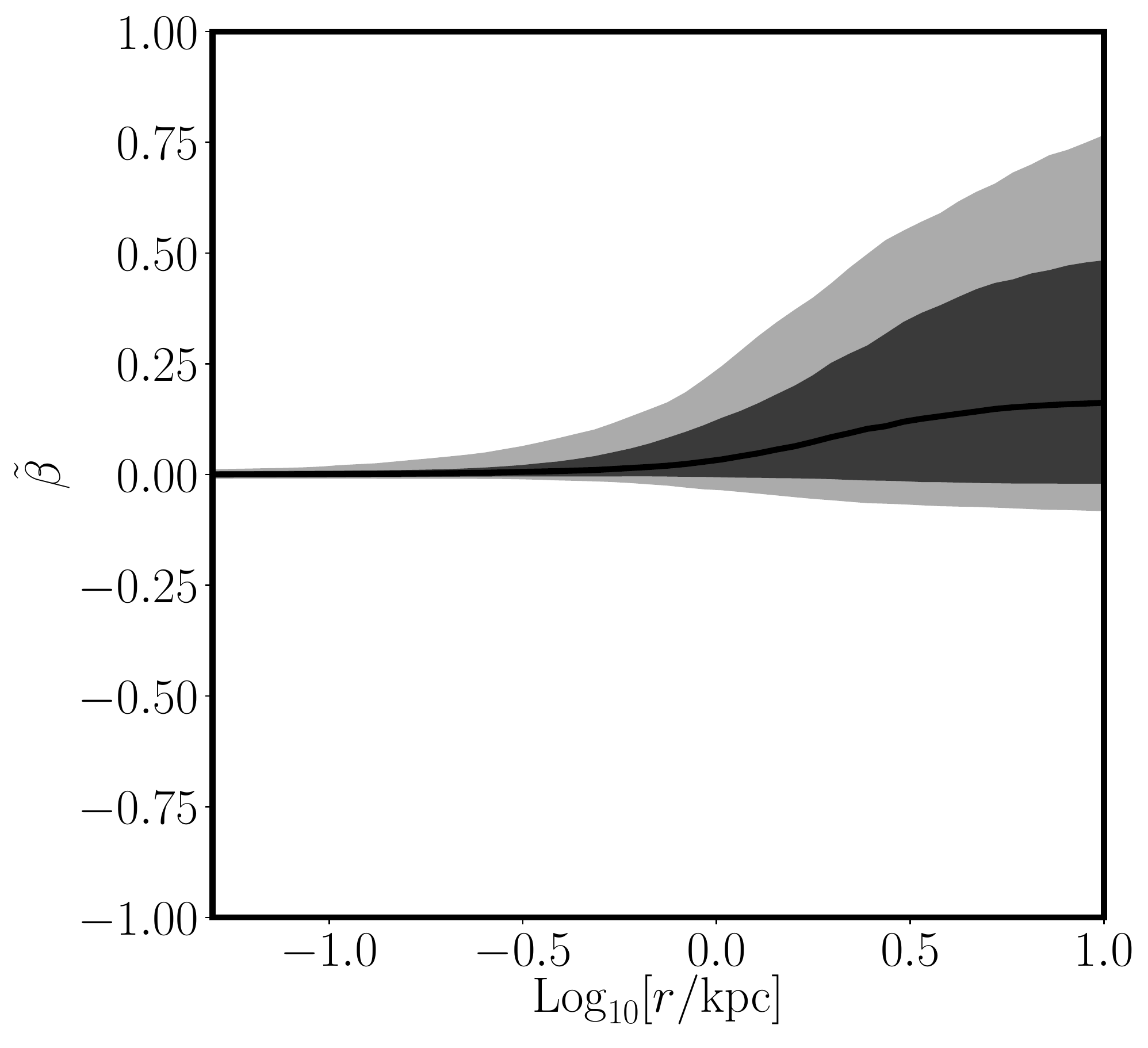}
     \hspace{-1mm}\includegraphics[angle=0,width=0.43\hsize]{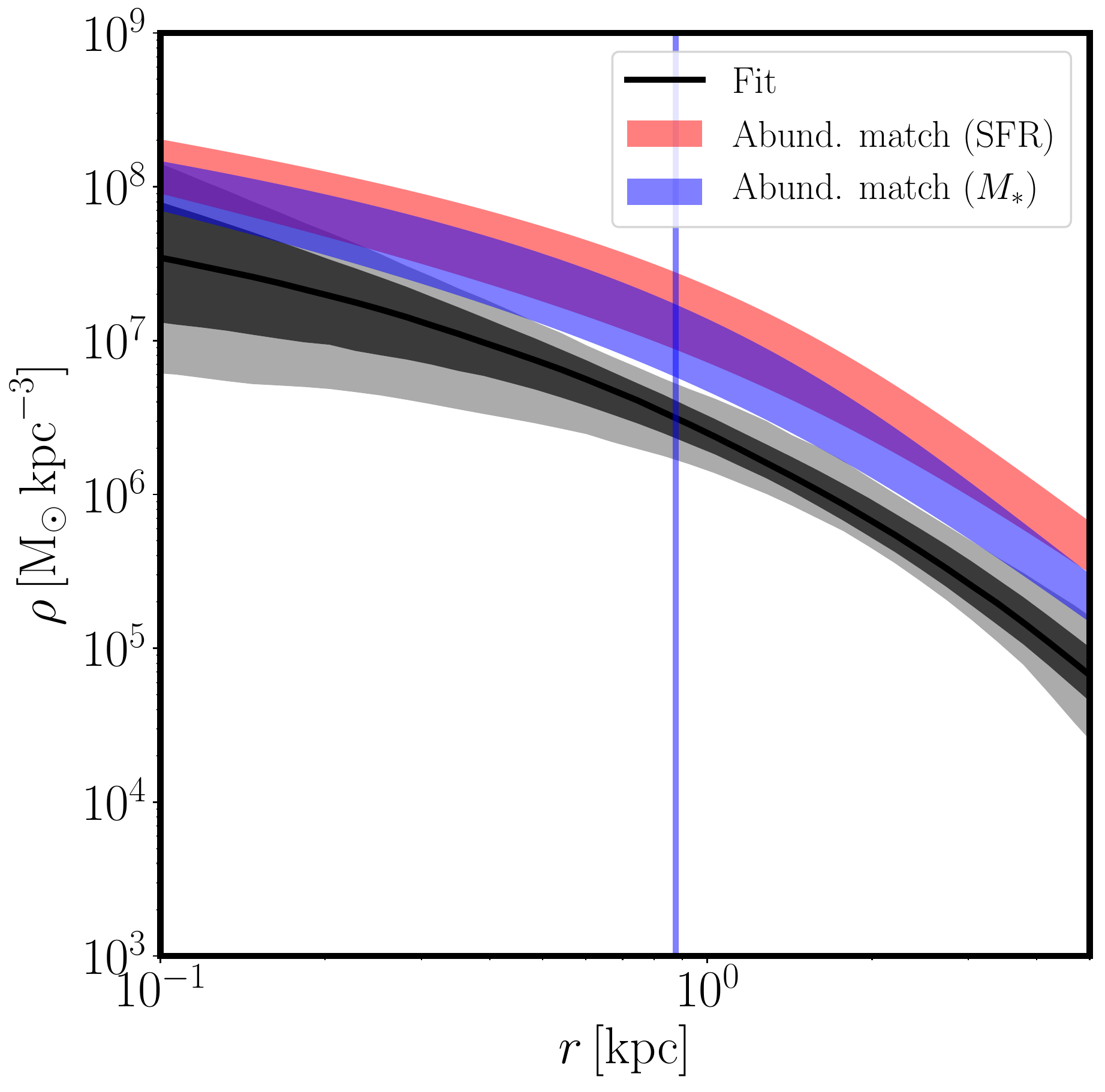}
      \caption{{\bf Top left:} Surface brightness profile for And XXI. The blue points show photometric data from deep Subaru imaging. The best-fit from {\sc GravSPhere} is shown as a solid black line, with the gray shaded regions showing the 68\% and and 95\% confidence intervals. {\bf Top right:} The velocity dispersion profile for And~XXI. Each bin contains an effective ${\sim}25$ stars weighted by membership probability. Again, the best fit and confidence intervals from \GravSphere\ are shown in black/grey. {\bf Bottom left:} The symmetrised radial anisotropy profile for And~XXI. {\bf Bottom right:} The final dark matter density profile inferred by \GravSphere. Notice that it is consistent with both a cusp and core within \GravSphere's 95\% confidence intervals (light grey). However, it is less dense at all radii than expected from abundance matching in $\Lambda$CDM, whether abundace matching with the stellar mass (blue) or the mean star formation rate (red). In all panels, the blue vertical line shows the half-light radius for And~XXI determined from PAndAS photometry \citep{martin16c}.}
  \label{fig:GSprof}
  \end{center}
\end{figure*}

We now apply {\sc binulator}+\GravSphere\ to And~XXI, with the goal of constraining its dark matter density profile. We take our surface brightness profile from our Subaru Suprime-cam imaging, which covers And~XXI out to 4 effective radii (this agrees within quoted uncertainties with that derived from PAndAS photometry \citep{martin16c}), and our dispersion profile is constructed from our 77 spectroscopically identified members. To generate this profile, the probabilities of our likely members are summed to give an ``effective'' number of members,

\begin{equation}
N_{\rm eff}=\sum\limits^{N_{\rm mem}}_{i=0} P_{\rm mem, i}.
\end{equation}

For our sample, $N_{\rm eff}=50$. We bin the data radially from the centre of And~XXI in two bins, using an effective 25 stars per bin. We show the $\Sigma_*(R)$ and $\sigma_{\rm LOS}(R)$ profiles in fig.~\ref{fig:GSprof}. The vertical blue line represents the half-light radius determined from the PAndAS imaging. We model And~XXI under the assumption that it is a spherical, non-rotating system, and we use the stellar mass from \citet{mcconnachie12} of $M_*=7.6\times10^5\,{\rm M}_\odot$, assuming an error of 25\%.

Using the above, we model the density profile of And~XXI and show the result in the lower left panel of fig.~\ref{fig:GSprof}. We measure a central dark matter density within 150\,pc of $\rho_{\rm DM}({\rm 150 pc})=2.6^{+2.4}_{-1.5} \times 10^7 \,{\rm M_\odot\,kpc^{-3}}$ at 68\% confidence. The data are consistent with a cusp or core within $2\sigma$ (light gray shading). Overlaid on the Figure are two density profiles determined from abundance matching in $\Lambda$CDM, using the stellar mass (blue) and the mean star formation rate (red), as in \citet{read19b} (a more detailed description of how these are derived is presented in \S\ref{sec:lowdenhalo}, below). For the latter, we use the star formation history for And~XXI from \citet{weisz19a}. Notice that in both cases, our determination of the dark matter density profile for And~XXI is lower {\it at all radii} than expectations from abundance matching in $\Lambda$CDM by a factor of ${\sim}3-5$.

In the following section, we discuss this low density profile in the broader $\Lambda$CDM context, and investigate whether dark matter heating from star formation or tidal processes could explain this result.

\section{Discussion}

\subsection{A low density halo from star formation}\label{sec:lowdenhalo}

Star formation can lower the central density of a dark matter halo, gradually turning a dark cusp into a core within ${\sim}$ the half light radius of the stars, $R_{1/2}$ \citep{navarro96b,read05,pontzen12,read16,read19a}. The star formation history of And~XXI has been measured from shallow HST data (reaching $\sim 1$ magnitude below the red clump) by \citet{weisz19a}. It formed $\sim50\%$ of its stars prior to 8.3\,Gyr ago, and $90\%$ prior to 5.8\,Gyr ago. There is no sign of any star formation in the last $2-3$\,Gyrs. Could this extended star formation be enough to lower the central density of And~XXI? 

To answer this question, we first need to estimate a pre-infall halo mass, $M_{200}$ for And~XXI. For this we use the abundance matching machinery from \citet{read19b}. Classic abundance matching estimates $M_{200}$ statistically from the stellar mass, $M_*$. For And~XXI, using the abundance matching relation from \citet{read17a} (consistent with \citealt{behroozi13}), this yields $M_{200} = 1.1 \pm 0.5 \times 10^9\,{\rm M}_\odot$. However, for satellite galaxies like And~XXI that quench on infall to a larger host, $M_*$ becomes a poor proxy for the pre-infall $M_{200}$ \citep[e.g.][]{ural15,tomozeiu16,read19b}. \citet{read19b} argue that using instead the mean star formation rate, $\langle {\rm SFR} \rangle$, evaluated over the period during which the satellite was forming stars, solves this `quenching problem'. They show empirically for Milky Way satellites that $\langle {\rm SFR} \rangle$ correlates much better with $M_{200}$ than does $M_*$. If we abundance match And~XXI using instead its $\langle {\rm SFR} \rangle = 2.1 \pm 0.5 \times 10^{-4} {\rm M}_\odot/{\rm yr}$ (calculated from And~XXI's star formation history as in \citealt{read19b}), we obtain a higher $M_{200} = 2.7 \pm 1.1 \times 10^9\,{\rm M}_\odot$.

Next, we must calculate whether or not there has been sufficient star formation within And~XXI to cause its central density to be lowered significantly. For this, we use the {\sc coreNFW} profile calibrated on the simulations from \citealt{read16}. \citet{read16} determine the amount of core formation based on the total `star formation time', defined as in \citet{read19b} as $t_{\rm SF} = M_* / \langle {\rm SFR} \rangle$. For And~XXI, this gives $t_{\rm SF} = 3.5$\,Gyrs. The amount of coring is then given by $n = {\rm tanh}(0.04 t_{\rm SF} / t_{\rm dyn})$, where $t_{\rm dyn}$ is the dynamical time at the scale radius, $r_s$ (equation \ref{eqn:MNFW}; and see \citealt{read16}). (Recall that $n = 0$ corresponds to complete core formation, while $n=1$ corresponds to a $r^{-1}$ density cusp.)

Assuming the \citet{dutton14} $M_{200}-c_{200}$ relation for $\Lambda$CDM, and its scatter, we use the above \coreNFW\ profile to estimate the range of theoretical expectations for And~XXI's density profile in $\Lambda$CDM. This is shown in Figure \ref{fig:GSprof}, bottom right panel. The blue band shows the expected range assuming $M_{200}$ derived from abundance matching with $M_*$; the red band shows the same using abundance matching with $\langle {\rm SFR} \rangle$. In both cases, the width of the band incorporates uncertainties in $M_{200}$ and the expected $2-\sigma$ scatter in $c_{200}$. Notice that in all models, the central density has been slightly lowered by star formation, but not enough to produce a flat core (we find that our models span the range $n = 0.2-0.5$). Further lowering of the inner density could be caused by late minor mergers, a new mechanism for dark matter heating reported recently in \citet{orkney21}. However, this still cannot explain the low density we find for $R > R_{1/2}$.

The end result of this exercise is that And~XXI's density is lower than expected for isolated halos in $\Lambda$CDM by a factor of ${\sim}3-5$. This is true at better than 95\% confidence at the half light radius ($R_{1/2} = 0.875$\,kpc) where the density is best-constrained. And, it remains true when marginalising over the expected $2-\sigma$ scatter in $c_{200}$ in $\Lambda$CDM, and when accounting for dark matter heating lowering the central dark matter density. We conclude, therefore, that And~XXI's density at all radii is lower than expected for isolated halos in $\Lambda$CDM. We discuss next whether this can be explained by tides.

\subsection{A low density halo from tidal processes?}

In the discussion above, we have assumed that And~XXI is unaffected by tides. Tidal stripping, and even more so tidal shocking, will act to lower And~XXI's density over time \citep[e.g.][]{gnedin99,read06a,read06b,amorisco19a}. If we assume that significant tidal stripping (losing 90-99\% of its original mass) has taken place, we can use the relations for tidal evolution from \citet{penarrubia10b} to assess the impact on And~XXI's density. These tidal tracks can be implemented for either cuspy or cored dark matter halos, and the effects for these are very different. If And~XXI possess an NFW cusp ($\gamma=1$), then tides have  almost no effect on its central density. Even assuming just 5\% of the original halo mass remains, the cusp survives and keeps the central density high. If instead we assume a small amount of cusp weakening due to baryonic processes (c.f. discussion above), then for $\gamma = 0.5$, and removal of 95\% of the original halo, we can find central density values that are consistent with our measurements for And~XXI within their 68\% confidence intervals. 

However, tidal shocking can be much more efficient than tidal stripping. The effect is maximised if And~XXI is on a plunging orbit and starts out with low density, either due to some inner cusp-core transformation, or due to it inhabiting a low concentration halo (see discussion in \S\ref{sec:intro} and \citealt{amorisco19a}).

The above results imply that we can explain And~XXI's low density through a combination of dark matter heating and tidal processes, with the dominant effect coming from tides. But, how likely is it that And~XXI is on an orbit that has allowed it to experience extreme tidal stripping or shocking? Both tidal scenarios require a highly radial orbit, with a small pericentre (<20\,kpc; \citealt{read06b}). And~XXI is currently far from its host, at a 3D distance of $D_{\rm M31}=145^{+11}_{-6}\,{\rm kpc}$ \citep{weisz19b}. Without proper motions, it is difficult for us to place meaningful constraints on the current orbit of And~XXI to determine whether it has recently passed close to M31. Even with proper motions, its current orbit is not necessarily a robust indicator of its past close interactions \citep[e.g.][]{lux10,genina21}. 
\par
The current light profile for And~XXI shows no obvious signs of tidal stripping in the outskirts, but this does not necessarily imply that tidal stripping of the stellar component has not taken place  \citep[e.g.][]{read06b,penarrubia09,ural15,genina21}. Furthermore, significant tidal shocking can lower the density of And~XXI if it is on a sufficiently plunging orbit, without any tidal stripping of stars taking place \citep{read06b, amorisco19a}. The unusual dynamics of And~XXI reported in \S~\ref{sect:kin} could imply that the system is not in dynamical equilibrium, but this $2\sigma$ finding is not adequate to provide unambiguous evidence of tidal stripping. With future data and modelling, better constraints can be placed on the orbit of And~XXI and on its faint stellar outskirts. This will help us understand whether tidal effects have acted to lower the central density of this dwarf galaxy, or whether its low density points to physics beyond $\Lambda$CDM.

\subsection{Comparison to predictions from modified Newtonian dynamics}

Finally, we consider other theories which may also explain And XXI's low central density and other properties. In \citet{mcgaugh13} , they calculate the velocity dispersions for Andromeda dSphs in the modified Newtonian dynamics (MOND) framework. As dwarf galaxies would be free from dark matter in this framework, one may expect to see a lower central mass or density than predicted by $\Lambda$CDM. Indeed, MOND has been shown to nicely reproduce the low density cores seen in isolated low surface brightness galaxies (e.g. \citealt{sanders02,famaey12}). In general, the MOND predictions of \citet{mcgaugh13} for  M31 dwarf spheroidals show a good consistency with the measured  velocity dispersions presented in both \citet{tollerud12} and \citet{collins13} (though note that some of the Milky Way dwarfs are more problematic for MOND; \citealt[e.g.][]{angus14,read19a}). Here, we compare our updated dispersion for And~XXI with these prior predictions.

\citet{mcgaugh13} give two estimates for the velocity dispersion of M31 dwarfs. The first is the isolated case, and the second is for dwarf galaxies embedded in an external field (the external-field effect, EFE). For satellite galaxies, it is not clear that the isolated approximation is valid, and for And~XXI in particular, they recommend using the EFE calculations. They predict a velocity dispersion of $\sigma_{\rm EFE}=3.7^{+1.5}_{-1.1}\kms$, where the presented value uses an assumption of the stellar mass to light ratio of $[M/L_*]=2\,{\rm M_\odot/L_\odot}$, and the uncertainties use $[M/L_*]=1~{\rm and}~4\,{\rm M_\odot/L_\odot}$. The uncertainty on the predicted dispersion reflects this range of M/L they considered to be plausible. This value is lower than our measured value of $\sigma_v=6.1^{+1.0}_{-0.9}\kms$, but it is within $1\sigma$. As such, the MOND prediction still agrees with our new data.

\section{Conclusions}

We have presented results from our chemodynamical study of the low mass M31 satellite, And~XXI. We have identified 77 probable stellar members in this galaxy, and our key findings are as follows:

\begin{itemize}
    \item We measure a systemic velocity  for And XXI of $v_r=-363.4\pm1.0\kms$ and a velocity dispersion of $\sigma_v=6.1^{+1.0}_{-0.9}\kms$, consistent with the findings of \citet{collins13} from a much smaller sample. 
    \item We find that the stars in the outskirts of And~XXI have a far lower velocity dispersion than those within the half-light radius. We also see that the systemic velocity and velocity dispersion vary with radius, which could indicate that it is not in dynamical equilibrium. 
    \item From those stars with high enough $S/N$, we additionally measure the metallicity distribution function of And~XXI. We measure a mean metallicity of ${\rm [Fe/H]}=-1.7\pm0.1$~dex. This is perfectly consistent with the stellar mass-metallicity relation for low mass galaxies \citep{kirby13a}. We are unable to measure the metallicity spread in And~XXI, but find it to be lower than 0.5 dex at 99\% confidence. 
    \item We model the dark matter density profile of And~XXI using an updated version of \GravSphere, with the main improvement being a complete reworking of its data binning routines into a separate code: the {\sc binulator} (\S\ref{sec:binulator}). This fits a generalised Gaussian PDF to each bin to estimate its mean, variance and kurtosis, and their uncertainties. This has the advantages that: 1) the distribution function can be readily convolved with the error PDF of each star, survey selection functions, binary star velocity PDFs, and similar; and 2) the method returns a robust estimate of the mean, variance and kurtosis even in the limit of a very small number of tracer stars. We tested this new method on mock data in Appendix \ref{app:mocks}, showing that it provides an unbiased estimate of the underlying density profile even for ${\sim}100$ tracer stars.
    \item Using {\sc binulator}+{\sc GravSphere}, we find a central dark matter density for And~XXI of $\rho_{\rm DM}({\rm 150 pc})=2.6_{-1.5}^{+2.4} \times 10^7 \,{\rm M_\odot\,kpc^{-3}}$ at 68\% confidence, and a density at two half light radii of $\rho_{\rm DM}({\rm 1.75 kpc})=0.9_{-0.2}^{+0.3} \times 10^6 \,{\rm M_\odot\,kpc^{-3}}$ at 68\% confidence. We cannot distinguish between a cusped or cored profile, however the density at all radii is a factor ${\sim}3-5$ lower than the densities expected from abundance matching in $\Lambda$CDM. We show that this cannot be explained by `dark matter heating' since And~XXI had too little star formation to significantly lower its inner dark matter density, while dark matter heating only acts on the profile inside the half light radius. However, And~XXI's low density can be accommodated within $\Lambda$CDM if it experienced extreme tidal stripping (losing $>95\%$ of its mass), or if it inhabits a low concentration halo on a plunging orbit that experienced repeated tidal shocks. Future work establishing the orbit of And~XXI would help us understand the origin of its low inner density and thereby improve our understanding of the nature of dark matter.
    \item When comparing our measured velocity dispersion for And~XXI with expectations from MOND, we find it to be consistent with predictions by \citet{mcgaugh13} within $1\sigma$. Given this object sits within the low acceleration regime, it will certainly be of interest for future work.
\end{itemize}

\section*{Data Availability}

All raw DEIMOS spectra are available via the \href{https://www2.keck.hawaii.edu/koa/public/koa.php}{Keck archive}. An electronic table with reduced properties (coordinates, magnitudes, $S/N$, velocities and metallicities) for all stars will be provided on the journal website. Fully reduced 1D spectra will be made available on reasonable request to the lead author as these are not hosted on the Keck archive. The data used in the \GravSphere\ modelling are available (with the code) at \href{https://github.com/justinread/gravsphere}{https://github.com/justinread/gravsphere}.

\section*{Acknowledgments}

We thank the referee for their helpful comments, which have improved the quality of this manuscript.

RI and NM acknowledge funding from the European Research Council (ERC) under the European Unions Horizon 2020 research and innovation programme (grant agreement No. 834148).

The data presented herein were obtained at the W. M. Keck Observatory, which is operated as a scientific partnership among the California Institute of Technology, the University of California and the National Aeronautics and Space Administration. The Observatory was made possible by the generous financial support of the W. M. Keck Foundation. Based in part on data collected at Subaru Telescope, which is operated by the National Astronomical Observatory of Japan.

The authors wish to recognize and acknowledge the very significant cultural role and reverence that the summit of Maunakea has always had within the indigenous Hawaiian community.  We are most fortunate to have the opportunity to conduct observations from this mountain.




\bibliographystyle{mnras}
\bibliography{michelle} 




\appendix

\section{Testing {\sc binulator}+{\sc GravSphere} on mock data}\label{app:mocks}

In \S\ref{sec:gravsphere}, we introduced an updated version of the \GravSphere\ code \citep{read17b,read18a} with a new reworked binning method, the {\sc binulator}. In this Appendix, we present tests of the new code on mock data drawn from the Gaia Challenge\footnote{ \href{http://astrowiki.ph.surrey.ac.uk/dokuwiki/}{http://astrowiki.ph.surrey.ac.uk/dokuwiki/}} spherical and triaxial suite \citep{read21}. We focus on two mocks that were particularly challenging for the previous version of \GravSphere: PlumCuspOm and PlumCoreOm. The former is a Plummer sphere embedded in a spherical, cuspy dark matter halo; the latter is the same embedded in a dark matter halo with a central constant density core. In both cases, the velocity dispersion profile is isotropic at the centre and maximally radially anisotropic at large radii. Both mocks are described in detail in \citet{read21}. Here, we model both assuming Gaussian velocity errors on each star of 2\,km/s for 100, 1,000 and 10,000 randomly sampled kinematic tracers. We assume that the photometric light profile is in all cases well-sampled with 10,000 stars, similarly to the situation with real data for nearby dwarf spheroidal galaxies \citep[e.g.][]{read19a}.

\begin{figure}
    \centering
    \includegraphics[width=0.47\textwidth]{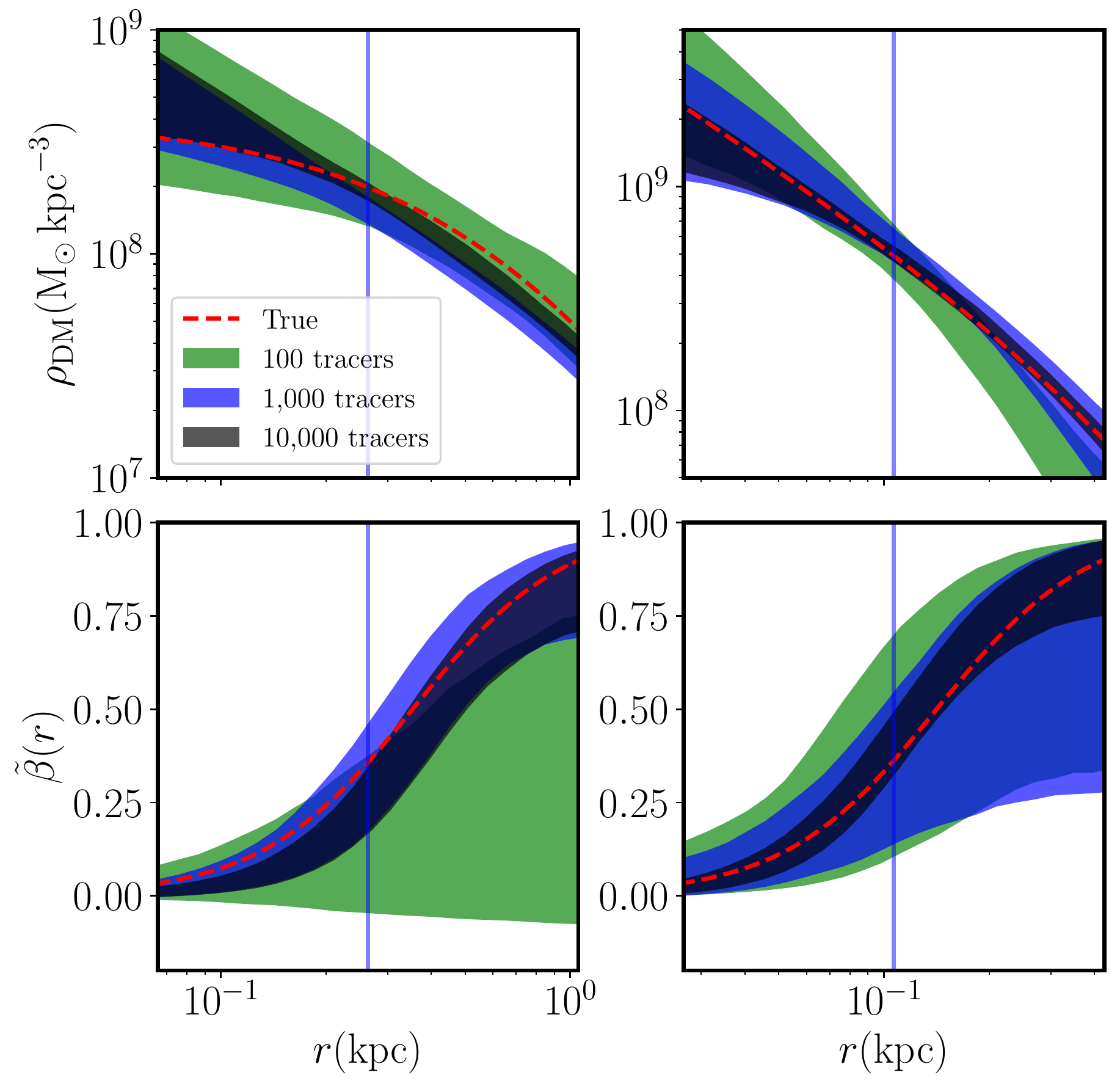}
    \caption{Testing the updated \GravSphere\ code and its new {\sc binulator} module on spherical mock data from the Gaia Challenge suite. The left panels show results for the PlumCoreOm mock; the right panels for the PlumCuspOm mock. The top panels show the recovery of the density profile for 100 tracers (green), 1,000 tracers (blue) and 10,000 tracers (black), where the width of the bands mark the 95\% confidence intervals in each case. The red dashed lines show the true solutions. The bottom panels show the same for the recovery of the symmetrised velocity anisotropy profile. Notice that, even with 100 tracers, we obtain an unbiased recovery of the radial density profile. For the PlumCuspOm mock, we are also able to weakly detect the radial anisotropy at large radii. As the sampling increases, the confidence intervals narrow around the correct solution.}
    \label{fig:GCmock}
\end{figure}

Previously, \GravSphere\ struggled on these mocks. For both PlumCoreOm and PlumCuspOm with 1,000 tracers, it returned a dark matter density that was systematically low for $R > R_{1/2}$, shifting, however, to $R > 4 R_{1/2}$ for 10,000 tracers. The velocity anisotropy was well-recovered in all cases, except PlumCoreOm for which 1,000 tracers was insufficient to detect the radial anisotropy at large radii \citep{read21}. In \citet{gregory19a}, \GravSphere\ was tested on mock data for fewer tracers (20, 100 and 500 tracer stars). There, a similar bias towards low densities at large radii was reported.

We show results applying the updated \GravSphere\ with its new {\sc binulator} binning module to the PlumCoreOm and PlumCuspOm mocks in Figure \ref{fig:GCmock}. In all cases, we obtain an unbiased recovery of both the dark matter density profile, $\rho_{\rm DM}(r)$, and the symmetrised stellar velocity anisotropy, $\tilde{\beta}(r)$. The code is no longer biased to low density at large radii. This owes to a combination of the improved data binning, the proper inclusion of the full error PDF for the virial shape parameters, and to our switch to using the {\sc coreNFW} profile (recall that the non-parametric power-law-in-bins profile used previously by default in \GravSphere\ biases models towards falling very steeply at large radii in the absence of data). Notice that now, even with just 100 tracers, we are able to detect that the PlumCuspOm is denser and falls more steeply than PlumCoreOm. For PlumCuspOm, we are also able to weakly detect that the velocity distribution is radially anisotropic at large radii. As we move to 1,000 and 10,000 tracers, we obtain an increasingly high-fidelity recovery of both the density profile and the velocity anisotropy.

\bsp	
\label{lastpage}
\end{document}